\documentclass[12pt]{article}

\usepackage[scale=.8]{geometry}
\usepackage[utf8]{inputenc}
\usepackage{booktabs}
\usepackage[titletoc]{appendix}

\usepackage{cite}
\usepackage{amsfonts,amsmath,amssymb,amsthm,mathtools}
\usepackage{bbm}
\usepackage{stackengine}\stackMath
\usepackage{xspace}
\usepackage{enumitem}
\usepackage{verbatim}
\usepackage{adjustbox}

\usepackage{graphicx}
\usepackage[usenames, dvipsnames]{xcolor}
\usepackage{tikz}
\usetikzlibrary{calc,shapes,matrix,arrows,calc,decorations.markings}
\usepackage{tikz-cd}
\usepackage{rotating}

\tikzset{commutative diagrams/.cd, arrow style=tikz, diagrams={>=latex}}

\usepackage{hyperref}

\makeatletter
\newtheorem*{rep@theorem}{\rep@title}
\newcommand{\newreptheorem}[2]{%
\newenvironment{rep#1}[1]{%
 \def\rep@title{#2 \ref{##1}}%
 \begin{rep@theorem}}%
 {\end{rep@theorem}}}
\makeatother

\newreptheorem{lemma}{Lemma}

\newreptheorem{conj}{Conjecture}
\newtheorem{claim}{Claim}[section]
\newreptheorem{claim}{Claim}

\theoremstyle{definition}

\theoremstyle{remark}
\newtheorem*{rem*}{Remark}

\newcommand{\fsu}{\mathfrak{su}}
\newcommand{\fso}{\mathfrak{so}}
\newcommand{\fsp}{\mathfrak{sp}}
\newcommand{\fg}{\mathfrak{g}}

\newcommand{\fe}{\mathfrak{e}}

\newcommand{\ord}{\text{ord}\xspace}

\newcommand*\xoverline[2][0.75]{
    \sbox{\myboxA}{$\m@th#2$}
    \setbox\myboxB\null
    \ht\myboxB=\ht\myboxA
    \dp\myboxB=\dp\myboxA
    \wd\myboxB=#1\wd\myboxA
    \sbox\myboxB{$\m@th\overline{\copy\myboxB}$}
    \setlength\mylenA{\the\wd\myboxA}
    \addtolength\mylenA{-\the\wd\myboxB}
    \ifdim\wd\myboxB<\wd\myboxA
       \rlap{\hskip 0.5\mylenA\usebox\myboxB}{\usebox\myboxA}%
    \else
        \hskip -0.5\mylenA\rlap{\usebox\myboxA}{\hskip 0.5\mylenA\usebox\myboxB}%
    \fi}
\makeatother

\begin{document}
 
~\vspace{2cm}
\begin{center}
{\huge\bfseries The Frozen Phase  of\\[10pt] Heterotic F-theory Duality }\\[10mm]

Paul-Konstantin Oehlmann$^{\,a,}$\footnote{\href{mailto:p.oehlmann@northeastern.edu}{p.oehlmann@northeastern.edu}}, Fabian Ruehle$^{\,a,b,c,}$\footnote{\href{mailto:f.ruehle@northeastern.edu}{f.ruehle@northeastern.edu}}, Benjamin Sung$^{\,d,}$\footnote{\href{mailto:bsung@ucsb.edu}{bsung@ucsb.edu}} \\[10mm]
\bigskip
{
	{\it ${}^{\text{a}}$ Department of Physics, Northeastern University, Boston, MA 02115, USA}\\[.5em]
	{\it ${}^{\text{b}}$ Department of Mathematics, Northeastern University, Boston, MA 02115, USA}\\[.5em]
	{\it ${}^{\text{c}}$ NSF Institute for Artificial Intelligence and Fundamental Interactions, Boston, MA, USA}\\[.5em]
    {\it ${}^{\text{d}}$ Department of Mathematics, University of California, Santa Barbara, CA 93106, USA }\\[.5em]
}
\end{center}
\setcounter{footnote}{0} 
\bigskip\bigskip

\begin{abstract}
\noindent 
We study the duality between the $\textit{Spin}(32)/\mathbbm{Z}_2$ heterotic string without vector structure and F-theory with frozen singularities. We give a complete description in theories with $6$d $\mathcal{N}=(1,0)$ supersymmetry and identify the duals of $\textit{Spin}(32)/\mathbbm{Z}_2$-instantons on ADE singularities without vector structure in the frozen phase of F-theory using an ansatz introduced by Bhardwaj, Morrison, Tachikawa, and Tomasiello. As a consequence, we obtain a strongly coupled description of orbifold phases of type I string theory without vector structure, substantially expanding the list of known examples of $6$d F-theory compactifications with frozen singularities. Supergravity theories can be \textit{fused} from these instanton theories, in a way that commutes with switching off vector structure, which we use to  propose new consistency checks via neutral hypermultiplet counting. Finally, we describe various Higgsings of this duality, and comment on constraints on higher form symmetries. 
\end{abstract}

\clearpage
\renewcommand{\baselinestretch}{0.6}\normalsize
\tableofcontents
\renewcommand{\baselinestretch}{1.}\normalsize
\clearpage

\section{Introduction}
String theory is the leading candidate for an ultraviolet (UV) complete theory of quantum gravity. In recent years, however, efforts have been dedicated towards understanding general principles of quantum gravity, independent of string theory. This collective effort is often referred to as the swampland program \cite{Ooguri:2006in} (for some recent reviews, see \cite{Grana:2021zvf,vanBeest:2021lhn,Agmon:2022thq})), and aims to study properties of seemingly consistent theories beyond the current string landscape. Progress in this direction is thus heavily guided by our knowledge of explicit constructions of string theory vacua.

F-theory is by far the leading systematic ansatz for the enumeration of theories and for the description of the massless content at low energies (see \cite{Weigand:2018rez} for a recent review). There is an explicit dictionary relating F-theory compactifications and the geometry of an elliptically fibered Calabi-Yau manifold, and details of the physics can be often extracted directly from the geometry. As a consequence, there is a rich interplay between new conjectures in the swampland program and concrete realizations of such proposals in explicit F-theory compactifications.

Nevertheless, there are aspects of F-theory compactifications which are not understood directly from the compactification geometry. Famously, compactifications to $4$ dimensions requires specifying $G_4$-flux, and additional data such as the location of spacetime-filling $D3$-branes often complicates the analysis of the effective physics. On the other hand, there are subtle non-geometric effects which appear even in higher dimensions; an M-theory compactification on an ADE singularity $\mathbbm{C}^2 /\Gamma$ comes with the choice of a discrete flux
\[
\int_{\mathbbm{C}^2 /\Gamma} C_3  = \frac{n}{d}
\]
for the $C_3$-field. Dualizing to F-theory potentially leads to a new class of supersymmetric defects, and~\cite{Tachikawa:2015wka} demonstrated that this must be a strongly coupled version of the familiar orientifold $7$-plane, but with positive Ramond-Ramond (RR) charge.

The positively charged orientifold $7$-plane ($O7^+$) is a supersymmetric defect which appears already in type IIB string theory and carries the same charge as a stack of $8$ $D7$-branes on top of a negatively charged orientifold $(O7^-)$. As a consequence, these two $7$-brane stacks cannot be distinguished by the $SL(2,\mathbbm{Z})$-monodromy in F-theory alone. The only effect of the $O7^+$-plane then, is that the corresponding singularity in the F-theory geometry cannot be deformed in the physical theory. This, however, cannot be deduced directly from the geometry alone. As a consequence of this subtle distinction, there has been little progress in constructing F-theory compactifications with frozen singularities and in analyzing their effective physics.

In fact, most of the advancements in studying such compactifications have only been made in $8$~dimensions with $16$ supercharges. 8D $\mathcal{N}=1$ supergravity vacua admit (at least) three disconnected moduli components, that have rank $20, 12$ and $4$ gauge algebra. The latter two components
were first studied in~\cite{Witten:1997bs} under the guise of the $\textit{Spin}(32)/\mathbbm{Z}_2$ heterotic string on $T^2$ without vector structure, which can be dualized to the CHL string\cite{Chaudhuri:1995fk,Font:2021uyw}, obtained from a $\mathbbm{Z}_2$ involution of the $E_8 \times E_8$ gauge group. Their F-theory realization must admit an $O7^+$ plane to implement the rank $8$ gauge group reduction and were first described in~\cite{Witten:1997bs}. Taken together, F-theory gives a complete description of all $8$d $\mathcal{N}=1$ supergravity theories, in beautiful agreement with predictions from the swampland program \cite{Bedroya:2021fbu}.

A first step in incorporating $O7^+$-planes in $6$d $\mathcal{N}= 1 $ F-theory compactifications was undertaken in~\cite{Bhardwaj:2018jgp}, and applications towards constructing new $6$d $\mathcal{N}=(1,0)$ gauge \cite{Bhardwaj:2019hhd} and  supergravity theories were discussed in~\cite{Morrison:2023hqx}. On the other hand, it is important to note that~\cite{Bhardwaj:2018jgp} gives only {\it necessary} conditions for {\it possibly} consistent constructions of frozen $6$d F-theory vacua, guided by gauge, gravitational, and mixed anomaly cancellations\footnote{In recent literature, type IIB compactifications with $O7^+$ planes have been explored in the context of SCFTs in 6D and 5D
in \cite{Bhardwaj:2019hhd} and \cite{Kim:2023qwh,Hayashi:2023boy,Kim:2024vci} respectively.  

}. Thus, in order to make progress in this direction, it is essential to expand the list of known examples, and to study such vacua through different duality frames whenever possible.

In this paper, we substantially expand the list of known examples by giving a complete construction of $6$d $\mathcal{N}=(1,0)$ string theory vacua in the frozen phase of the duality between the $\textit{Spin}(32)/\mathbbm{Z}_2$ heterotic string and F-theory. More precisely, we study various nonperturbative limits of the $\textit{Spin}(32)/\mathbbm{Z}_2$ heterotic string involving small instantons on $ADE$-singularities without vector structure, and we describe such vacua through F-theory with $O7^+$-planes using the rules derived in~\cite{Bhardwaj:2018jgp}. As a consequence, we conjecture new consistency conditions for constructing frozen F-theory backgrounds, and we give a description for $\textit{Spin}(32)/\mathbbm{Z}_2$ heterotic compactifications without vector structure near strongly coupled points in moduli space. Our main results can be summarized as follows:
\begin{repclaim}{clm:duals}
There exists frozen F-theory compactifications dual to the $\textit{Spin}(32)/\mathbbm{Z}_2$ heterotic string without vector structure in the limit that small $\textit{Spin}(32)/\mathbbm{Z}_2$-instantons collide with an ADE singularity.
\end{repclaim}
\begin{repclaim}{clm:neutrals}
Let $\mathcal{T}_1$ and $\mathcal{T}_2$ be two 6D $\textit{Spin}(32)/\mathbbm{Z}_2$ heterotic compactifications related by turning off vector structure. Then they have the same number of unlocalized neutral hypermultiplets.
\end{repclaim}
This paper is structured as follows: In Section~\ref{sec:heteroticwovector} we construct the orbifold theories of $\textit{Spin}(32)/\mathbbm{Z}_2$ instantons, with and without vector structure.  In Section~\ref{sec:fheorydual} we discuss the frozen F-theory duals of the former theories. In Section~\ref{sec:frozengravity} we use those NS5 brane orbifolds as building blocks for SUGRA theories. In Section~\ref{sec:1FormSyms}, we discuss the discrete center 1-form gauge symmetry sector in the frozen phase. We present our conclusions in Section~\ref{sec:Conclusions}. Appendix~\ref{app:ToricData} contains details of the geometric construction used in Section~\ref{sec:frozengravity}.

\section{Heterotic Strings Without Vector Structure}\label{sec:heteroticwovector}
In order to motivate and review the following discussion in six dimensions, we will review the eight-dimensional case first.
 
Our starting point will be the $G=\textit{Spin}(32)/\mathbbm{Z}_2$ string compactified on a torus. Moreover, we can switch on a non-trivial $G$-connection which
leads to an adjoint Higgsing of the 10D gauge group. This straight-forward compactification leads to a rank$(G)$ + $U(1)^4=20$ gauge group in eight dimensions. The fact that $G$ is non-simply connected also allows to switch on a non-trivial holonomy along the $T^2$ that lifts to $G$ but not to its $\textit{Spin}(32)$ cover. Indeed, as the $\mathbbm{Z}_2$ quotient, projects out the vector (and co-spinor) representation, the holonomy is said to have no vector structure \cite{Witten:1997bs}.  

The $\mathbbm{Z}_2$ holonomy factor acts non-trivially on the $\fsu_2 \in \fso_{32}$ leaving an $\fsp_8$ residual gauge algebra factor. By this discrete choice, we have effectively reduced the gauge group rank by eight, such that we are in a rank 12 component of 8D SUGRA vacua, where also the CHL heterotic string lives \cite{Lerche:1997rr}.

The same vacuum can be also constructed in type IIB string theory. Here we can start with a torus compactification as well but in order to reduce the amount of supersymmetry we require an orientifold. The $T^2/\mathbbm{Z}_2$ has four fixed points, which gives the space the topology of a $\mathbbm{P}^1$. At each fixed point resides an $O7$ plane.

For the $O7$ planes, we have a choice of a sign of the Ramond-Ramond charges,
\begin{align}
Q_{RR}(O7^\pm) = \pm 4 \, \,\qquad \text{ with }\qquad Q_{RR}(D7)=1 \, . 
\end{align}
Both planes back-react with the type IIB axio-dilaton $\tau=C_0 + i / g_\text{IIB}$.  The axion back-reaction of an $O7^-$ plane may be cancelled by adding 4 $D7$ branes which yields an $\fso_8$ gauge algebra. Putting four more $D7$ branes on top produces an $\fso_{16}$ gauge algebra. This brane stack has the same axio-dilaton backreaction as a single $O7^+$ brane, which does not carry a gauge symmetry. Hence at this level, when trading the $O7^-$ brane stack as 
\begin{align}
(8D7 + O7^-) \quad\leftrightarrow\quad O7^+ \, ,
\end{align}
we have achieved a gauge group rank reduction by $8$, which puts us in the IIB dual of the rank$(G)$=12 moduli space of the 8D Heterotic $\textit{Spin}(32)/\mathbbm{Z}_2$ theory without vector structure or the CHL vacua. 

Indeed, we can move in the moduli space of this 8D theory and place eight more $D7$ branes on top of the $O7^-$ stack. For 16 $D7$ branes we can then trade 
\begin{align}
(16 D7 + O7^-) \quad\rightarrow\quad (8D7 + O7^+) 
\end{align}
which exactly results in changing the gauge algebra as $\fso_{32} \rightarrow \fsp_8$.

The 8D gauge theory is a great starting point to discuss the lower six-dimensional gauge theories that live on heterotic NS5 branes. From the perspective of the 6D theory, we have decoupled gravity but the heterotic 8D gauge theory stays as a flavor symmetry of a little string theory. 

In the following, we discuss the 6D gauge theory on $\textit{Spin}(32)/\mathbbm{Z}_2$ heterotic NS5 brane both with and without vector structure. In Section~\ref{sec:fheorydual} we  
geometrize those theories to type IIB/F-theory.

\subsection[Review: \texorpdfstring{$SO(32)$}{SO(32)} NS5 branes without vector structure]{Review: \texorpdfstring{$\boldsymbol{SO(32)}$}{SO(32)} NS5 branes without vector structure}
One significant advancement in our understanding of string theory lies in the application of dualities to study highly non-perturbative and singular points in the moduli space of string compactifications. A particular example is the $SO(32)$ heterotic string; there is a limit in which the curvature of the gauge bundle is localized at a point on the compact background, corresponding to an instanton shrinking to zero size. The dynamics governing the small instanton is by now fairly well understood~\cite{Witten:1995gx}; when the small instanton is localized at a smooth point, there is an additional $SU(2)$ gauge symmetry localized at the point. The case of small instantons supported at ADE singularities was further explored in~\cite{Intriligator:1997kq,Blum:1997mm}, and the relevant physics was given a geometric interpretation from F-theory in~\cite{Aspinwall:1997ye}, see also~\cite{Ludeling:2014oba,Cvetic:2018xaq}. In this section, we will briefly review the results for ADE singularities, and we will defer the discussion of their F-theoretic interpretation to Section~\ref{sec:ftheoryns5}.

The main considerations of~\cite{Intriligator:1997kq,Blum:1997mm} can be summarized as follows: we first consider a configuration of $K$ instantons localized at an ADE singularity on an ALE space $X$. To specify the configuration, we will need to specify the holonomies at infinity surrounding the singularity; we will assume that these are set to zero for now. The moduli space of instantons on $X$ can then be given a hyperk\"ahler quotient construction as the Higgs branch of a supersymmetric gauge theory. At a smooth point, this gauge theory can be realized physically as the worldvolume theory of precisely $K$ type I $D5$-branes in the background of $32$ $D9$-branes. 

One of the main insights of~\cite{Intriligator:1997kq,Blum:1997mm} was that at a singular point, the gauge theory arising in the hyperk\"ahler quotient construction cannot be realized physically as the worldvolume theory of type I $D5$-branes. There is an anomaly, but a minor modification of the physics results in a consistent type I construction. On the other hand, this modification does not describe the moduli space of instantons directly; instead, it realizes a different branch related by exchanging $29$ hypermultiplets for $1$ tensor multiplet. Summarizing, small $SO(32)$ instantons at an ADE singularity on a blown down ALE space, $X$, admit a Higgs branch described by the moduli space of instantons on a resolution of $X$, but they also admit what today is called a tensor branch described by type I $D5$-branes localized at the singularity. 

In the subsequent sections, we will primarily be concerned with the case of non-trivial holonomy, specified by a map
\[
\pi \colon \Gamma_G \rightarrow \textit{Spin}(32)/\mathbbm{Z}_2
\]
where $\Gamma_G$ is the fundamental group of a $3$-sphere surrounding the ADE singularity. There is an alternative $\mathbbm{Z}_2$-quotient of $\textit{Spin}(32)$ which is isomorphic to the group $SO(32)$; there is an element which maps to the identity in $SO(32)$, but maps to a non-trivial element $w$ in the usual $\textit{Spin}(32) / \mathbbm{Z}_2$ gauge group of the heterotic string. We will consider the case that $\pi$ obeys the group relations of $\Gamma_G$ up to this element $w$, known as the case without vector structure, and we will derive the effective gauge groups on the Coulomb branch via a weakly coupled type I orientifold analysis. This was first done in~\cite{Douglas:1996sw}, and we will extend this analysis to the cases of $D$ and $E$ singularities. 
 
\subsection{Examples} 
In this section, we use the type I dual frame to compute the gauge groups appearing on the tensor branch of small $SO(32)$ instantons on ADE singularities. Specifically, we examine in detail the cases of types $A$ and $D$ singularities with and without vector structure. In the following we will exemplify the algorithm in the case of $A_n$ and $D_4$ type singularities. 

\subsubsection[\texorpdfstring{$A_n$}{An} singularities]{\texorpdfstring{$\boldsymbol{A_n}$}{An} singularities}\label{sec:Ansing}
In \cite{Douglas:1996sw,Intriligator:1997kq}, the gauge groups describing the effective physics of small $SO(32)$ instantons on $A_n$-singularities were derived, by appealing to the dual type I description of $D5$-branes subject to an orbifold action. In this section, we review the relevant results and compute the gauge groups in both the cases with and without vector structure.

We begin with type IIB string theory with an orientifold action $\Omega$, and the orbifold group $A_n$ with generator $r$. The orientifold action is localized along the spacetime filling $O9^-$ plane, which projects out the $D7$-branes. There are 32 $D9$-branes required by tadpole cancellation, and we will focus on the action of these group elements on the $D5$-brane sector. Our goal is to find the effective gauge symmetries in the $D5$-brane sector, represented by matrices $U$ satisfying the conditions:
\begin{align}\label{eqn:gauge}
\begin{split}
U &= -\gamma(\Omega)U^{T} \gamma(\Omega)^{-1} \, , \\
U &= \gamma(r) U \gamma(r)^{-1} \, , \\
\end{split}
\end{align}
where $\gamma(\Omega)$ and $\gamma(r)$ are the unitary matrices representing the action of the orientifold and orbifold groups on the Chan-Paton labels of the $D5$-branes. In addition, we will assume that $\gamma(r)$ takes the following form:
\[
\gamma(r) = \begin{pmatrix}
V_0 & & & & \\ 
& V_1 & & & \\
&& \ddots & &\\
& & & V_{n-2} & \\
& & & & V_{n-1}
\end{pmatrix}, \quad
V_i = \begin{pmatrix}
\xi^i & & \\
& \ddots & \\
& & \xi^i
\end{pmatrix} \, ,
\]
where $\xi$ is a primitive $n$th root of unity.

These matrices satisfy a number of constraints imposed by the group relations, which take the following form:
\begin{align}\label{eqn:Anorientifold}
\begin{split}
\gamma(\Omega) &= - \gamma(\Omega)^T \, , \\
\gamma(r) \gamma(\Omega) \gamma(r)^T &= \chi(r,\Omega) \gamma(\Omega) \, , \\
\gamma(r)^n &= \chi(r) \mathbbm{1} \, . 
\end{split}
\end{align}
Critically, the unitary matrices need to satisfy the group relations only up to a phase factor, represented by the elements $\chi(r,\Omega)$ and $\chi(r)$ above.
The latter one can be removed, by scaling the $\gamma(r)$ matrices, up to an overall root of unity $\xi$. Rescaling $\gamma(r)$ by $\xi$ in \eqref{eqn:Anorientifold} amounts to rescaling $\chi(r,\Omega)$ to $\xi^2 \chi(r,\Omega)$ which then leaves two choices to rescale this phase factor as (see~\cite{Klein:2000qw} for more details) 
\begin{align*}
\chi(r,\Omega) = \left\{\begin{array}{ll}
1 &\text{($n$ odd) }\\
1,~\xi &\text{($n$ even)}
\end{array}\right.
\end{align*}
In the case with $\chi(g,\Omega) = 1$ after applying the second equation in~\eqref{eqn:gauge}, we find that $U$ must be constrained to take the following form:
\begin{align*}
U = \begin{pmatrix}
U_0 & & & & \\ 
& U_1 & & & \\
&& \ddots & &\\
& & & U_{n-2} & \\
& & & & U_{n-1}
\end{pmatrix} \, ,
\end{align*}
where $U_1,\ldots,  U_n \in SU(v_1), \ldots , SU(v_n)$. The first equation of~\eqref{eqn:gauge} then imposes
\begin{align*}
&U_0 \in \mathfrak{sp}(v_0), \ U_1 \in \mathfrak{su}(v_1), \ldots, U_{n/2-1} \in \mathfrak{su}(v_{n/2-1}),\ U_{n/2} \in \mathfrak{sp}(V_{n/2})\quad\text{for $n$ even} \\
&U_0 \in \mathfrak{sp}(v_0), \ U_1 \in \mathfrak{su}(v_1), \ldots, U_{n/2-1} \in \mathfrak{su}(v_{(n-1)/2}),\ \phantom{U_{n/2} \in \mathfrak{sp}(V_{n/2})~\,} \text{for $n$ odd} 
\end{align*}
In both cases, the degrees of freedom of all the other matrices $U_i$ with $ i > n/2$ are fixed by $U_0, \ldots, U_{n/2}$. Finally, taking into account the hypermultiplet sector, we find the following gauge theory chains
\[
\begin{tikzcd}[row sep=0mm]
\mathfrak{sp}(v_1) \arrow[r,dash]& \mathfrak{su}(v_2) \arrow[r,dash] & \ldots \arrow[r,dash] & \mathfrak{su}(v_{n-1}) \arrow[r,dash]& \mathfrak{sp}(v_{n/2}) \, ,\\
\mathfrak{sp}(v_1) \arrow[r,dash]& \mathfrak{su}(v_2) \arrow[r,dash] & \ldots \arrow[r,dash] & \mathfrak{su}(v_{(n-1)/2})  \, .
\end{tikzcd}
\]
for $n$ even and odd, respectively. In the case without vector structure, which only exists for $n$ even, we find the following gauge theory chain:
\[
\begin{tikzcd}
\mathfrak{su}(v_1) \arrow[r,dash]& \mathfrak{su}(v_2) \arrow[r,dash] & \ldots \arrow[r,dash] & \mathfrak{su}(v_{n-1}) \arrow[r,dash]& \mathfrak{su}(v_n)
\end{tikzcd}
\]

\subsubsection[\texorpdfstring{$D_4$}{D4} singularities]{\texorpdfstring{$\boldsymbol{D_4}$}{D4} singularities}\label{sec:Dnsing}
In this section, we carefully examine the cases of small $SO(32)$-instantons on a $D_4$-singularity both with and without vector structure, which has not appeared in the literature to the best of our knowledge.

Our analysis proceeds identically to the previous section; the presentation for the orbifold group $D_4$ takes the following form: 
\[
D_4 = \langle r,s \vert r^4 = 1, \ s^4 = 1, \ (rs)^4 = 1 \rangle \, .
\]
Compared to equation~\eqref{eqn:gauge}, the matrices $U$ describing the effective gauge symmetries in the $D5$-brane sector satisfy one additional condition corresponding to the generator $s$:
\begin{align}\label{eqn:gauged}
\begin{split}
U &= -\gamma(\Omega)U^{T} \gamma(\Omega)^{-1}  \, ,\\
U &= \gamma(r) U \gamma(r)^{-1} \, , \\
U &= \gamma(s) U \gamma(s)^{-1} \, . \\
\end{split}
\end{align}
We will assume that the Chan-Paton matrices corresponding to the generators $r$ and $s$ take the following forms~\cite[Appendix A]{Blum:1997mm}:
\begin{align*}
\gamma(r) &= \begin{pmatrix}
V_0 & & & & \\ 
& V_1 & & & \\
&& V_2 & &\\
& & & V_3 & \\
& & & & V_4
\end{pmatrix}, \quad V_2 = \begin{pmatrix}
i & 0 & & &  \\
0 & -i & & &  \\
& & \ddots & & & \\
& & & i & 0 \\
& & & 0 & -i
\end{pmatrix} \, ,\\[12pt]
V_0 &= V_4 = \mathbbm{1}\, , \quad V_1 = V_3 = - \mathbbm{1}\, , \\[12pt]
\gamma(s) &= \begin{pmatrix}
W_0 & & & & \\ 
& W_1 & & & \\
&& W_2 & &\\
& & & W_3 & \\
& & & & W_4
\end{pmatrix}, \quad
W_2 = \begin{pmatrix}
0 & 1 & & &  \\
-1 & 0 & & &  \\
& & \ddots & & & \\
& & & 0 & 1 \\
& & & -1 & 0
\end{pmatrix} \,,\\[12pt]
W_0 &= W_3 = \mathbbm{1}\, ,\quad W_1 = W_4 = - \mathbbm{1}\, .
\end{align*}
In order to the find general form for the matrices $U$, it suffices to compute the Chan-Paton matrix, $\gamma(\Omega)$, corresponding to the orientifold action. This must satisfy the following relations:
\begin{align}\label{eqn:orientifold}
\begin{split}
\gamma(\Omega) &= - \gamma(\Omega)^T  \, ,\\
\gamma(r) \gamma(\Omega) \gamma(r)^T &= \chi(r,\Omega) \gamma(\Omega) \, , \\
\gamma(s) \gamma(\Omega) \gamma(s)^T &= \chi(s, \Omega) \gamma(\Omega) \, ,\\
\end{split}
\end{align}
where $\chi(r,\Omega)$ and $\chi(s,\Omega)$ are phase factors. As in the above section, we refer to~\cite{Klein:2000qw} for a detailed discussion, and we simply note that there are two discrete choices corresponding to the cases with vector structure and without vector structure. 

\subsubsection*{With vector structure}
We recover the gauge algebras
\[
\mathfrak{sp}(v_0) \times \mathfrak{sp}(v_1) \times \mathfrak{so}(v_2) \times \mathfrak{sp}(v_3) \times \mathfrak{sp}(v_4)  \, ,
\]
by an orientifold analysis of the $D_4$-singularity. We consider the case:
\[
 \chi(r,\Omega) = 1\,, \quad \chi(s,\Omega) = 1 \,, 
\]
which corresponds to the case with vector structure.

After applying the second and third conditions in equation~\eqref{eqn:gauged}, the matrix $U$ must take the form
\[
U=
\begin{pmatrix}
U_0 &&&& \\
& U_1 &&& \\
& & U_2 && \\
&&& U_3 & \\
&&&&U_4 
\end{pmatrix} \, ,
\]
where $U_i \in \mathfrak{su}(v_i)$ for all $i$. The first equation of~\eqref{eqn:gauged} then demands that
\[
U_0 \in \mathfrak{sp}(v_0), \ U_1 \in \mathfrak{sp}(v_1), \ U_2 \in \mathfrak{so}(v_2), \ U_3 \in \mathfrak{sp}(v_3), \ U_4 \in \mathfrak{sp}(v_4) \, .
\]
Finally, an analysis of the orientifold action on the hypermultiplet sector leads to the quiver diagram
\[
\begin{tikzcd}
&\mathfrak{sp}(n_2) \arrow[d,dash] &  \\
\mathfrak{sp}(n_1) \arrow[r,dash]& \mathfrak{so}(n_3) \arrow[r,dash] \arrow[d,dash]& \mathfrak{sp}(n_4) \\
& \mathfrak{sp}(n_5)&
\end{tikzcd} \, ,
\]
in agreement with the results of~\cite{Blum:1997mm}. The full quiver, that includes the flavor and 2-form tensor charges is given in Table~\ref{fig:LSTQuivers}. 

\subsubsection*{Without vector structure}
We claim that the case without vector structure leads to the gauge algebras
\[
\mathfrak{su}(v_0) \times \mathfrak{su}(v_1) \times \mathfrak{sp}(v_2) \, ,
\]
where we take the phase factors to satisfy
\[
\chi(r,\Omega) = 1\, ,\quad  \chi(s,\Omega) = -1 \,. 
\]
The only difference from the previous subsection comes from the first equation in~\eqref{eqn:gauged}, which forces $\gamma(\Omega)$ to take a different form due to the choice of phase factors.

Imposing the conditions
\begin{equation}\label{eqn:so}
\gamma(r) U \gamma(r)^{-1} = U, \quad \gamma(s) U \gamma(s)^{-1} = U, \quad -\gamma(\Omega)U^T\gamma(\Omega)^{-1} = U \, .
\end{equation}
means that the $n \times n$ matrix $U$ has to be of the form
\[
U=
\begin{pmatrix}
U_0 &&&& \\
& U_1 &&& \\
& & U_2 && \\
&&& U_3 & \\
&&&& U_4 
\end{pmatrix} \, ,
\]
where $U_0 \in \mathfrak{su}(n_0), U_1 \in \mathfrak{su}(n_1), U_2 \in \mathfrak{so}(n_2)$. The degrees of freedom for $U_3$ and $U_4$ are completely fixed by $U_0$ and $U_1$, in analogy with the case of $A_n$ singularities without vector structure.

Finally, an analysis of the orientifold action on the hypermultiplet sector leads to the following quiver diagram
\[
\begin{tikzcd}
\mathfrak{su}(n_1) \arrow[r,dash] & \mathfrak{sp}(n_1) \arrow[r,dash] & \mathfrak{su}(n_2) \, .
\end{tikzcd}
\]

\subsection{Summary of results}
\label{sec:ResultSummary}
The above procedure can be readily applied to all types of ADE singularities. 
The residual exceptional discrete groups
 $\Gamma_{E_n}$ for $n=6,7,8$ are given by the binary tetrahedral, binary octahedral and binary icosahedral groups, respectively. Their respective presentations  are given by the generators 
\begin{align*}
E_6 &= \langle r,s\, \vert \,  (r s)^2 = s^3 = r^3=-1 \rangle \, ,\\
E_7 &= \langle r,s \,\vert \, (rs)^2 = s^3=r^4 = -1 \rangle \, ,\\
E_8 &= \langle r,s,t \, \vert \, (st)^2 = s^3 = t^5= 1 \rangle \, .
\end{align*}
We see that the presentation of $E_6$ and $E_7$ contains even powers of the generators, while the presentation of $E_8$ also has odd powers. This is important, since the general considerations for switching off vector structure (see Section~\ref{sec:Ansing}) demand that we can pick the phase $\chi(x,\Omega)=-1$, which is only possible if the powers are even. We thus find that one can freeze $E_6$ and $E_7$, but not $E_8$. 

The resulting frozen and unfrozen cases are summarized in Table~\ref{fig:LSTQuivers}. It is noteworthy that the frozen quivers have the shape of the respective ADE affine extended Dynkin diagram, but folded by an $\mathbbm{Z}_2$ automorphism. It is known that the binary tetrahedral, octahedral and icosahedral groups, also called the binary polyhedral groups, form a subgroup of the automorphism group of the Dynkin diagrams of $E_n$. From that perspective, $E_8$ singularities cannot be frozen since their respective (extended) Dynkin diagram does not admit any automorphisms.  Interestingly, the little string quivers in Table~\ref{fig:LSTQuivers} have also been found as T-duals of $E_8 \times E_8$ heterotic LSTs in \cite{DelZotto:2022xrh, Ahmed:2023lhj}, but with a $\mathfrak{u}_{16}$ flavor group (or a subgroup therefore) instead of $\fsp_8$. This suggests that similar quivers can be constructed in the unfrozen phase of F-theory.  It would be very interesting to elucidate this connection in future works. 
 
\begin{table}[ht!]
\begin{center}
\begin{tabular}{|c|c|c|}\hline
$G$ & Quiver w/ Vector Structure $\mathcal{K}^{UF}_M(G)$ & Quiver  w/o Vector Structure $\mathcal{K}^{F}_M(G)$ \\ \hline
$\emptyset$ & $[\fso_{32}] \overset{\fsp_M}{0}$ & $ [\fsp_{8}]  \overset{\fsp_{M}}{0}$ \\ \hline 
$A_1$ & $[\fso_{32}]  \overset{\fsp_{M+4}}{1} \overset{\fsp_{M}}{1} $ & $[\fsp_{8}]  \overset{\fsu_{2M}}{0} $ \\  \hline
$A_3$ & $[\fso_{32}] \overset{\fsp_{M+8}}{1} \overset{\fsu_{2M+8}}{2} \overset{\fsp_{M}}{1} $ & $[\fsp_{8}] \overset{\fsu_{2M+8}}{1} \, \, \overset{\fsu_{2M}}{1}   $ \\  \hline

\hspace{-0.1cm}$A_{2n-1}$ &  
$[\fso_{32}] \overset{\fsp_{M+4n}}{1} \overset{\fsu_{2M+8(n-1)}}{2} \,\overset{\fsu_{2M+8(n-2)}}{2} ... \overset{\fsu_{2M+8}}{2} \, \,  \overset{\fsp_{M}}{1} $
&   \hspace{-0.5cm}
$[\fsp_{8}] \hspace{-0.1cm}\overset{\fsu_{2M+8n}}{1} \overset{\fsu_{2M+8(n-1)}}{2} \overset{\fsu_{2M+8(n-2)}}{2} ...  \overset{\fsu_{2M+8}}{2} \, \,  \overset{\fsu_{2M}}{1} $ \hspace{-0.4cm}
   \\  \hline \hline
$D_4$ &$
[\fso_{32}] \overset{\fsp_{8+M}}{1} \, \,  \overset{ \displaystyle \overset{\fsp_M}{ 1}}{  \underset{ \displaystyle \overset{\fsp_M}{1}}{\overset{\fso_{16+4M}}{4}}}  \, \, \overset{\fsp_{M}}{1} 
$  & $      \, \,  \overset{ \displaystyle [\fsp_{8}] \overset{  \fsu_{M+8}}{ 2} \quad \, \, \,  }{ \underset{\displaystyle \, \, \overset{\fsu_M}{2}}{\, \, \overset{\fsp_{M}}{1}}}  \, \, $ \\ \hline   

\hspace{-0.1cm}$D_{2n+4}$  &
 
$\lbrack \fso_{32} \rbrack    \,     \overset{\fsp_{4n+M+8}}{1  } \,\overset{ \displaystyle \overset{\fsp_{4N+M}}{1} }{  \overset{\fso_{16N+4M+16}}{4  } }\,\overset{\fsp_{8n+2M }}{1  }    ...\overset{ \displaystyle \overset{\fsp_{M}}{1} }{\overset{\fso_{4M+16}}{4  } }   \,     \overset{\fsp_{M}}{1  }  $
 & 
 $ \hspace{-0.4cm}
 \overset{ \displaystyle [\fsp_{8}] \hspace{-0.2cm}  \overset{ \fsu_{M+8+4n}}{ 2} \quad \, \, \,  }{ \underset{ \displaystyle \, \,  \overset{  \fsu_{M+4n}}{2}}{\, \, \overset{\fsu_{2M+8n}}{2}}} \hspace{-0.7cm}  \overset{\fsu_{2M+8(n-1)}}{2}  \, \overset{\fsu_{2M+8(n-2)}}{ 2 } ...   \overset{\fsu_{2M+8}}{ 2\ } \,\overset{\fsp_{M }}{1} $\hspace{-0.4cm}
 
  \\ \hline \hline

$D_{5}$ & $[\fso_{32}] \, \, \overset{\fsp_{8+M}}{1} \, \,  \overset{ \displaystyle \overset{\fsp_{M}}{ 1}}{
\overset{\fso_{16+4M}}{4} 
} \, \, \overset{\fsp_{2M }}{1}\, \,  \overset{\fsu_{2M }}{2} \,  $  & 
$
  \, \,  \overset{ \displaystyle [\fsp_{8}] \overset{  \fsu_{M+8}}{ 2} \quad \, \, \,  }{ \underset{\displaystyle \, \, \overset{\fsu_M}{2}}{\, \, \overset{\fsu_{2M}}{1}}}  \, \, $ \\ \hline
 \hspace{-0.1cm}$D_{2n+5} $&
 $[\fso_{32}] \, \, \overset{\fsp_{8+M+4n}}{1} \, \,  \overset{ \displaystyle \overset{\fsp_{ M+4n}}{ 1}}{
\overset{\fso_{16+4M+16n}}{4} 
} ...    \overset{\fsp_{2M }}{1} \,  \overset{\fsu_{2M  }}{2} \,  $
 &
 $
   \hspace{-0.15cm}  \overset{ \displaystyle [\fsp_{8}] \hspace{-0.2cm} \overset{ \fsu_{M+8+4n}}{ 2}\quad \, \, \,  }{ \underset{ \displaystyle \, \overset{\fsu_{M+4n}}{2}}{\, \, \overset{\fsu_{2M+8n}}{2}}}  \hspace{-0.7cm}   \, \overset{\fsu_{2M+8(n-1)}}{2} \, \overset{\fsu_{2M+8(n-2)}}{ 2}  \hspace{-0.1cm}...  \hspace{-0.1cm}\overset{\fsu_{2M+8}}{ 2}  \,\overset{\fsu_{2m}}{1}   $ \hspace{-0.2cm}
 \\ \hline \hline 
 $E_6$ & $ 
 [\fso_{32}] \, {\overset{\mathfrak{\fsp}_{M+8}}{1}} \, {\overset{\fso_{4M+16}}{4}}\,  {\overset{\mathfrak{\fsp}_{3M }}{1}} \, {\overset{\mathfrak{\fsu}_{4M }}{2}} \, {\overset{\mathfrak{\fsu}_{2M}}{2}}
 $ &    $
  [\fsp_{8}]  \, {\overset{\mathfrak{\fsu}_{2M+16}}{2}} \,  {\overset{\mathfrak{\fsu}_{4M+16}}{2}} \, 
 {\overset{\mathfrak{\fsp}_{3 M +8 }}{1}}\, {\overset{\fso_{4 M+16}}{4}} \, {\overset{\mathfrak{\fsp}_{2 M}}{1}}$ 
 \\ \hline 
 \hline 
  
 $E_7$ & $    \hspace{-0.2cm}  
  [\fso_{32}]  \hspace{-0.2cm} {\overset{\mathfrak{\fsp}_{M+12}}{1}} \, {\overset{\fso_{4M+32}}{4}}\,  {\overset{\mathfrak{\fsp}_{3M+12}}{1}} \,   \overset{\displaystyle \overset{\fsp_{2M+4}}{1}}{{\overset{\fso_{8M+32}}{4}}} \, {\overset{\mathfrak{\fsp}_{3M+8}}{1}} \,  {\overset{\fso_{4M+16}}{4}}  \, \overset{\fsp_{M}}{1}  \hspace{-0.2cm}$ 
  & $
   [\fsp_{8}]  \, {\overset{\mathfrak{\fsu}_{2M+12}}{2}} \,  {\overset{\mathfrak{\fsu}_{4M+8}}{2}} \, {\overset{\mathfrak{\fsu}_{6M+4}}{2}} \, 
 {\overset{\mathfrak{\fsp}_{4M}}{1}}\, {\overset{\fso_{4M+8}}{4}}
$  
  \\ \hline 
\end{tabular}
\caption{\label{fig:LSTQuivers}The quivers $\mathcal{K}$ of $G$ orbifolds of $M$
$\textit{Spin}(32)/\mathbbm{Z}_2$ instantons with and without vector structure.  }
\end{center}
\end{table}

\section{F-theory duals}\label{sec:fheorydual}
\subsection{Unfrozen models}\label{sec:ftheoryns5}

In Section~\ref{sec:heteroticwovector}, we analyzed the tensor branch of small $SO(32)$-instantons at an ADE singularity without vector structure, by relying on their description via $D5$-branes in perturbative type I string theory. In particular, this description is accurate for large vacuum expectation values of the tensor multiplet, far out on the tensor branch of the interacting fixed point. On the other hand, the same physics can be described purely geometrically via F-theory~\cite{Aspinwall:1997ye}, which has the advantage that it also gives a description for the superconformal fixed point, as well as the transition to the Higgs branch. In this section, we will review aspects of F-theory compactifications to six dimensions, focusing on the F-theory duals to the $SO(32)$-heterotic string and small instanton transitions appearing in~\cite{Aspinwall:1997ye}.

In contrast to lower-dimensional compactifications, we can specify a $6$d F-theory model via an elliptically fibered Calabi-Yau threefold $X$ over a smooth complex surface $B$. We assume that there exists a section, in which case there exists birational transformations of $X$ to a Weierstrass equation
\begin{equation}\label{eq:weierstrass}
y^2 = x^3 + fx + g, \qquad \Delta= 4f^3 + 27g^2\,,
\end{equation}
with $f$ and $g$ sections of $-4K_B$ and $-6K_B$ respectively, and we denote by $-K_B$ the anticanonical divisor on $B$. Note that $B$ must be compact to yield a SUGRA theory; when it is not, the resulting theory is a QFT that flows to an LST or SCFT.
The general fiber is an elliptic curve which degenerates along a curve, $\Delta \subset B$, often called the discriminant locus. In general, $\Delta$ will be irreducible, but in physically interesting situations with multiple gauge factors and matter content, there is a decomposition $\Delta = \Delta_1 \cup \ldots \cup \Delta_n$ into smooth irreducible algebraic curves $\Delta_i$. The details of the singularities can be deduced by monodromy actions on the $1$-cycles of the torus fiber induced by loops around the components, $\Delta_i$, which correspond to elements in $SL(2,\mathbbm{Z})$.

Compactifications to six dimensions give $6$d $\mathcal{N} = (1,0)$ supergravity theories, which have been studied extensively in recent years~\cite{Kumar:2009ac,Kumar:2010ru,Morrison:2011mb,Morrison:2012np,Morrison:2012js,Kumar:2010am,Kumar:2009ae,Kumar:2009us,Park:2011wv,Monnier:2017oqd}. Theories in six dimensions are distinguished, in that they are constrained by gauge, gravitational, and mixed anomalies, which allow for a systematic study of their massless gauge algebras and matter content. Simultaneously, the $6$d $\mathcal{N} = (1,0)$ F-theory landscape is still rich enough to evade a systematic classification, which is consistent with the lack of classification results for elliptic Calabi-Yau threefolds. 

Finally, we note that there is a rich dictionary relating details of the $6$d $\mathcal{N} = (1,0)$ supergravity theory with singularities of the corresponding elliptic Calabi-Yau threefold. The irreducible codimension-one components $\Delta_i$ of the discriminant $\Delta \subset B$ determine the localized $7$-brane stacks which support the non-abelian gauge algebras. The intersection points $\Delta_i \cap \Delta_j$ host localized matter, corresponding to massless string states with ends on $\Delta_i$ and $\Delta_j$. The $SL(2,\mathbbm{Z})$ monodromies of the irreducible discriminant loci $\Delta_i$ correspond bijectively to the singular Kodaira fibers, and these together with the intersection data determine the non-abelian gauge algebras and massless matter content as deduced in~\cite[Tables 1, 2]{Grassi:2014zxa} and ~\cite{Grassi:2011hq}.

We will focus on a specific example, known as the Aspinwall-Gross model, which was first described in~\cite{Aspinwall:1996nk}, and admits a frozen variant described in~\cite{Bhardwaj:2018jgp}. We take the base $B = \mathbbm{F}_4$, the Hirzebruch surface with a unique curve $e$ of self-intersection $(-4)$, with Weierstrass polynomials
\begin{align}\label{eqn:weierstrass}
\begin{split}
f &= \frac{1}{3}e^2(-p_2^2 + 3 e^6 q^{24})\,,\qquad
g = \frac{1}{27}e^3 p_2 (-2 p_2^2 + 9 e^6 q^{24})\\[12pt]
\Delta &= e^{18} q^{48} ( - p_2 + 2e^3 q^{12} )(p_2 + 2e^3 q^{12})\,,
\end{split}
\end{align}
where $q$ denotes the fiber class, and $p_2$ is a generic polynomial of class $3e + 12q$. We note that a concrete tuning realizing such polynomials on $B$ can be found in~\cite{Aspinwall:1996nk}. From an analysis of the $6$d effective physics, we observe that $e$ has vanishing orders $(2,3,12)$ and hence corresponds to an $I_{12}^*$-singularity with an $\mathfrak{so}_{32}$-algebra, $q$ has vanishing orders $(0,0,48)$ and corresponds to an $I_{48}^\text{ns}$-singularity with an $\mathfrak{sp}_{24}$-algebra (if it was split, this would be an $\mathfrak{su}_{48}$ algebra), and there is a single bi-fundamental localized at the intersection. The precise details of the duality to the $SO(32)$ heterotic string can be found in~\cite{Aspinwall:1997ye}; we note here that there is the required $\mathfrak{so}(32)$ gauge symmetry, together with an $\mathfrak{sp}(24)$ gauge symmetry realized by $24$ small instantons coalesced at a smooth point.

One of the main results of~\cite{Aspinwall:1997ye} was the description of the F-theory dual to the limit where $24$ small instantons move on top of an ADE-singularity. This limit leads to a strongly interacting fixed point with extra massless tensor multiplets, and the physics far out on the tensor branch was described through type I string theory in Section~\ref{sec:heteroticwovector}. In the case of a $\mathbbm{Z}_2$-singularity, the physics on the tensor branch can be described via a $\mathbbm{Z}_2$ orbifold of type~I string theory. From Table~\ref{fig:LSTQuivers}, we see that there is a transition
\[
\begin{tikzcd}
\overset{\mathfrak{sp}_{24}}{0} \rar[mapsto]& \overset{\mathfrak{sp}_{24}}{1} &[-35pt]  \overset{\mathfrak{sp}_{20}}{1} 
\end{tikzcd}
\]
where the $\mathfrak{sp}_{24}$ gauge symmetry is enhanced to the product $\mathfrak{sp}_{24} \times \mathfrak{sp}_{20}$ together with a bi-fundamental hypermultiplet as the $24$ small instantons move from a smooth point to a $\mathbbm{Z}_2$-singularity. 

This phase transition, often called the small instanton transition, is famously described by an F-theory compactification in the limit where there exists a point $p$ in the base $B$ such that the orders of vanishing of the Weierstrass polynomials satisfy:
\[
(\ord_p(f), \ord_p(g), \ord_p(\Delta) ) \geq (4,6,12)
\]
The details of this transition can be found in~\cite{Apruzzi:2018oge}, and we note that this leads to an F-theory compactification on a Calabi-Yau Weierstarss model over the blowup of the base $B$ at the point $p$, with exceptional curve $C$. The orders of vanishing along $C$, which determine the corresponding Kodaira singular fiber, are given by
\[
\ord_{C}(f) = \ord_p(f) - 4, \quad \ord_{C}(g) = \ord_p(g) - 6, \quad \ord_{C}(\Delta) = \ord_p(\Delta) - 12
\]

For the Weierstrass model~\eqref{eqn:weierstrass}, we note that the locus $\{ p_2 = 0 \}$ intersects with the line $\{ q = 0\}$ at $3$ points. The collision of two of these points leads to a point with orders of vanishing $(4, 6, 52)$, and a blowup leads to an additional curve with orders of vanishing $(0, 0, 40)$. Thus, there is an $I_{40}^\text{ns}$-fiber over this exceptional curve $C$, leading precisely to the enhancement we derived above via an orientifold analysis.

\subsection{Frozen rules}\label{sec:f} 

In this section, we will review the necessary rules for specifying a frozen F-theory compactification, as first derived in~\cite{Bhardwaj:2018jgp}. In Section~\ref{sec:ftheoryns5}, we defined a $6$d F-theory compactification through the choice of a Weierstrass model over a base surface $B$. We will define a $6$d F-theory compactification in the frozen phase similarly, but with the additional choice of replacing a stack of $7$-branes with the monodromy of an $I_{n+4}^*$-fiber with a stack of $7$-branes consisting of an $O7^+$-plane and $n$ $D7$-branes. This has the same monodromy as an $I_{n+4}^*$ fiber and to distinguish the two situations, we will label such a configuration with $\widehat{I}_{n+4}^*$.

One of the key features in the frozen phase of F-theory is that in general, neighboring $7$-brane stacks must share a common gauge algebra. We note that there exist only necessary conditions for specifying a consistent frozen F-theory compactification via Green-Schwarz anomaly cancellation, but that these are not sufficient in general. We assume that the discriminant locus $\Delta$ contains irreducible components $\Delta_a$, and we denote the corresponding $SL(2,\mathbbm{Z})$-monodromies by $M_a$.

We will specify a $6$d F-theory compactification in the frozen phase with the following data:
\begin{enumerate}[label=(\alph*)]
\item
To each irreducible component $\Delta_a$ such that the monodromy $M_a$ is conjugate to $M_{I_n^*}$, we may assign an unfrozen/frozen $7$-brane, denoted by $I_n^*$/$\widehat{I}_n^*$ respectively. We denote the sum of irreducible divisors $D_a$ supporting a frozen $7$-brane by $F$.
\item
A collection of gauge algebras $\mathfrak{g}_i$, and embeddings $\rho_{i,a} \colon \mathfrak{g}_i \xhookrightarrow{} \mathfrak{l}_a$ such that $\bigoplus_i \rho_{i,a}(\mathfrak{g}_i) \subset \mathfrak{l}_a$. Here, $\mathfrak{l}_a$ denotes the naive gauge algebra associated with the Kodaira fiber of the irreducible component $\Delta_a$. 
\end{enumerate}
The associated {\it gauge divisor} is then defined by 
\[
S_i \coloneqq \sum\limits_a \mu_{i,a} o_{i,a} D_a
\]
where $\mu_{i,a} = 0$ if $\rho_{i,a} = 0$, and is $1$ if $\rho_{i,a}$ is non-trivial. The coefficient $o_{i,a}$ is known as the Dynkin index of the embedding $\rho_{i,a}$, see~\cite{MR47629} for a precise definition. 

As argued in~\cite{Bhardwaj:2018jgp}, the anomaly polynomial needs to be modified in the presence of $O7^+$-planes, such that it takes the form
\[
I_{GS}^8 = -\frac{1}{2} \left(-(K+F) \frac{p_1(T)}{4} + \sum\limits_i \Sigma_i\; \nu(F_i)\right)^2
\]
where $p_1(T)$ is the Pontryagin class of the tangent bundle of $B$ and $\nu(F_i)$ is the instanton number density of the field strength $F_i$ valued in $\mathfrak{g}_i$, i.e. the second Chern class. Indeed, when $F = 0$, this agrees with the anomaly polynomial from a conventional $6$d F-theory compactification, and this suggests simply replacing the usual matter assignment with the substitution
\[
K \mapsto K +F 
\]
which leads to additional consistency conditions. For the resulting frozen F-theory compactification to be consistent, we will also demand that the following conditions hold:

\begin{enumerate}
\item
$\Sigma_i \cdot \Sigma_i \in \mathbbm{Z}$, $\Sigma_i \cdot \Sigma_j \in \mathbbm{Z}_{\geq 0 }$ \text{ for all } $i,j$
\item
$(K+F) \cdot \Sigma_i + \Sigma_i \cdot \Sigma_i = -2$
\item
For each pair $(\mathfrak{g}_i,\Sigma_i)$, the total number of localized charged hypermultiplets must sum up to the total number of hypermultiplets of the corresponding representation, as given in~\cite[Table 3.1]{Bhardwaj:2018jgp}.
\item
$n_{H,\text{charged}} - n_V < 273$
\end{enumerate}
Briefly, we can summarize the first three conditions as imposing the cancellation of gauge anomalies, and the last as imposing the usual gravitational anomaly cancellation. For simplicity, we will only consider cases where all gauge divisors intersect transversally, hosting matter in the bi-fundamental representation, with the exception of $\mathfrak{so}-\mathfrak{sp}$ intersections which host a half bi-fundamental.

Finally, we illustrate a simple example corresponding to the frozen variant of the Aspinwall-Gross model summarized in Section~\ref{sec:ftheoryns5}. We make the replacement $I_{12}^* \longrightarrow \widehat{I}_{12}^*$ on the exceptional divisor $e$ and we define the following frozen and gauge divisors
\[
F = e, \quad \Sigma_1 = \frac{1}{2}e, \quad \Sigma_2 = 2q 
\]
where $\Sigma_1$ supports an $\mathfrak{sp}(8)$ gauge algebra, and $\Sigma_2$ supports an $\mathfrak{su}(24)$ gauge algebra. The canonical embedding $\mathfrak{su}(24) \xhookrightarrow{} \mathfrak{su}(48)$ is an embedding of index $2$, which contributes a factor of $2$ in the definition of $\Sigma_2$. Finally, a straightforward intersection calculation implies
\[
(K+F) \cdot \Sigma_1 = -1, \quad \Sigma_1 \cdot \Sigma_1 = -1, \quad (K+F) \cdot \Sigma_2 = -2, \quad \Sigma_2 \cdot \Sigma_2 = 0, \quad \Sigma_1 \cdot \Sigma_2 = 1\,,
\]
which, by anomaly cancellation, gives $1$ bifundamental of $\mathfrak{sp}(8) \times \mathfrak{su}(24)$ and $2$ antisymmetrics of $\mathfrak{su}(24)$. The gauge group appearing on the fiber class $q$ in this compact case is indeed different from the expected gauge group of transverse $D7$-branes intersecting an $O7^+$-plane.

\subsection{Summary of results}\label{sec:summary}
In Section~\ref{sec:f}, we reviewed a simple example from~\cite{Bhardwaj:2018jgp} describing the frozen analog of the Aspinwall-Gross model. The freezing mechanism in this case describes precisely the mechanism of turning off vector structure in the $SO(32)$ heterotic string. Our main goal in this section is to summarize our ansatz for realizing the frozen F-theory duals of small $SO(32)$ instantons at arbitrary ADE singularities, as first described in Section~\ref{sec:heteroticwovector}. We will comment on the general patterns realized by such constructions for general frozen F-theory compactifications, and we will defer the precise analysis of such examples to Section~\ref{sec:frozenexamples}.

Our main results can be summarized with the following claims:
\begin{claim}\label{clm:duals}
There exist frozen F-theory compactifications dual to the $SO(32)$ heterotic string without vector structure in the limit that small $SO(32)$-instantons collide with an ADE singularity.
\end{claim}
\begin{claim}\label{clm:neutrals}
Let $\mathcal{T}_1$ and $\mathcal{T}_2$ be two 6D $SO(32)$ heterotic compactifications related by turning off vector structure. Then they have the same number of unlocalized neutral hypermultiplets.
\end{claim}
\noindent
We begin by summarizing the constructions leading to Claim~\ref{clm:duals}. Consider an F-theory compactification on a tuned Calabi-Yau elliptic fibration $X$ over $\mathbbm{F}_4$, with an $\mathfrak{so}_{32}$ gauge symmetry on the unique curve of self-intersection $(-4)$. As described in Section~\ref{sec:ftheoryns5}, one of the main results of~\cite{Aspinwall:1997ye} was the derivation of the physics describing small instantons on arbitrary ADE singularities purely in the context of F-theory. In Section~\ref{sec:heteroticwovector}, we derived the effective physics in the case without vector structure. On the other hand, a fiber-wise application of heterotic/F-theory duality implies that the F-theory dual to the $SO(32)$ heterotic string on a $K3$ orbifold without vector structure should be described by F-theory with a frozen $7$-brane localized on the $(-4)$-curve.

We will substantiate this claim by explicitly constructing the corresponding frozen F-theory compactifications for almost all of the ADE singularities. We will illustrate the procedure briefly for the case of the $A_1$-singularity, which already appeared in~\cite{Bhardwaj:2018jgp}. In this case, turning off vector structure induces the following map:
\[
\begin{tikzcd}
 \overset{\mathfrak{sp}_{24}}{1} & [ -35pt] \overset{\mathfrak{sp}_{20}}{1}\rar[mapsto] &\overset{\mathfrak{su}_{24}}{0} 
\end{tikzcd}
\]

where we have indicated the relevant gauge algebras together with the self-intersection of their curves of support. In particular, one begins with two $(-1)$-curves in the original F-theory geometry, contracts one of them, and then constructs a frozen embedding in the resulting Weierstrass model over the resulting base. This procedure can be generalized to arbitrary ADE singularities, and in general, we will contract as many curves as necessary to achieve the desired number of tensor multiplets derived in the model without vector structure.

Finally, we will briefly comment on Claim~\ref{clm:neutrals}. We expect that the above procedure does not change the number of complex structure deformations of the total space of the Weierstrass model, which leave the relevant singularities invariant. In particular, we expect that the number of unlocalized neutral hypermultiplets is invariant after turning off vector structure. We prove that this is true in all the models that we construct, and we summarize this in Table~\ref{fig:neutralsvector}.
\begin{table}[t]
\centering
\begin{tabular}{c|c|cccccccc|cccc|c}
\toprule
 &Singularities&$A_0$ &$A_1$& $A_2$ & $A_3$ & $A_4$ & $A_5$ & $A_6$ & $A_7$&  $D_4$ & $D_5$& $D_6$& $D_7$ &$E_6$  \\
\midrule
w/ vector&Tensors &0& 1& 1& 2& 2 &3 &3& 4& 4 & 4 & 5& 5 & 4  \\
structure&$H_\text{neutral}$&20& 19& 18& 17& 16& 15& 14& 13& 16& 15 & 14 & 13 & 14\\ 
\midrule
w/o vector&Tensors &0& 0& $\emptyset$& 1&  $\emptyset$ &2 & $\emptyset$&3&  2 & 2 & 3 & 3 & 4\\
structure&$H_\text{neutral}$&20& 19&  $\emptyset$& 17&  $\emptyset$& 15&  $\emptyset$& 13&  16 & 15 & 14 & 13 & 14 \\ 
\bottomrule
\end{tabular}
\caption{\label{fig:neutralsvector}  Tensor multiplets and neutral hypermultiplets in the $6$d supergravity models with and without vector structure. For convenience, we have specialized to the Aspinwall-Gross model with all $24$ small instantons localized at the same point.}
\end{table}

\subsection{Frozen models}\label{sec:frozenexamples}

In Section~\ref{sec:summary}, we derived the effective physics of small $SO(32)$ instantons on orbifold points through a weakly coupled type I orientifold analysis. As described in Section~\ref{sec:ftheoryns5}, the same physics can be described through F-theory as in~\cite{Aspinwall:1997ye} via heterotic F-theory duality. In this section, we will realize the results of Section~\ref{sec:summary} without vector structure in the frozen phase of F-theory.

\subsubsection[\texorpdfstring{$A_n$}{An} singularities]{\texorpdfstring{$\boldsymbol{A_n}$}{An} singularities}
In this section, we begin with the case of small $SO(32)$ instantons on $A_n$ singularities. The main results were summarized in the first column of Table~\ref{fig:LSTQuivers}; for $M$ instantons on an $A_{2n-1}$ singularity, the effective physics describing the tensor branch of the theory is given by the following quiver
\[
\text{Unfrozen:}~\quad[\fso_{32}] \overset{\fsp_{M+4n}}{1} \, \overset{\fsu_{2M+8(n-1)}}{2} \,\overset{\fsu_{2M+8(n-2)}}{2} \ldots \overset{\fsu_{2M+8}}{2} \, \,  \overset{\fsp_{M}}{1} 
\]
Our goal in this section is to realize the corresponding frozen quiver
\[
\text{Frozen:}~~~\quad[\fsp_{8}] \overset{\fsu_{2M+8n}}{1} \, \overset{\fsu_{2M+8(n-1)}}{2} \, \overset{\fsu_{2M+8(n-2)}}{2} \ldots \overset{\fsu_{2M+8}}{2} \, \,  \overset{\fsu_{2M}}{1} 
\]
geometrically via the rules of Section~\ref{sec:f}. In the following we discuss the various frozen instanton models, to be already part of a compact base for concreteness. In such cases we choose the number of instantons on the single theory to be already $24$. More general cases are discussed in Section~\ref{sec:frozengravity}.

We begin with the simplest examples, which already appeared in~\cite{Bhardwaj:2018jgp} and were summarized in Section~\ref{sec:f}. Consider the Aspinwall-Gross model, given by an F-theory compactification on the Hirzebruch surface $\mathbbm{F}_4$ with Weierstrass polynomials
\begin{align*}
f = e^2 \widetilde{f}, \qquad g = e^3 \widetilde{g},  \qquad \Delta = e^{18} q^{48} \widetilde{\Delta}\,.
\end{align*}
The relevant physics of this model can be summarized by the diagram
\[
\begin{tikzcd}
\text{Unfrozen:}~\quad\left [\fso_{32}\right ] &[-35pt]\overset{\mathfrak{sp}_{24}}{0} &[-10pt] [F] \arrow[dash]{r} & [-20pt] D_1
\end{tikzcd}
\]
where the divisor $F$ corresponds to the curve $\{ e = 0\}$ with self-intersection $(-4)$, and $D_1$ corresponds to a curve in the fiber $\{q  = 0 \}$ with self-intersection $(0)$. In particular, this describes a global $6$d model on the tensor branch of $24$ small instantons on a smooth point.
As described in~\cite{Bhardwaj:2018jgp} and Section~\ref{sec:f}, we claim that the model obtained by turning off vector structure is described by the same geometry, but with the following assignment of gauge divisors:
\[
\begin{tikzcd}[row sep=0pt]
\text{Frozen:}~\quad\left [\fsp_{8}\right ] &[-35pt]\overset{\mathfrak{sp}_{12}}{0} &[-10pt] [\frac{1}{2}F] \arrow[dash]{r} & [-20pt] 2D_1
\end{tikzcd}
\]
Here, we have flipped the $I_{12}^*$ to its frozen variant $\widehat{I}_{12}^*$ supported on the divisor class $\frac{1}{2}F$. In addition, we note that there exists an embedding $\mathfrak{sp}(12) \xhookrightarrow{} \mathfrak{su}(48)$ of index $2$, leading to the divisor class $2D_1$ supporting the gauge algebra $\mathfrak{sp}(12)$.

Next, we discuss $24$ small instantons on top of an $A_1$-singularity. As described in~\cite{Aspinwall:1997ye}, this realizes the following gauge theory chain in the unfrozen phase of F-theory :
\[
\begin{tikzcd}
\text{Unfrozen:}\quad~\left [\fso_{32}\right ] &[-35pt]\overset{\mathfrak{sp}_{24}}{1}&[-35pt]\overset{\mathfrak{sp}_{20}}{1} &[-10pt][ F] \arrow[dash]{r} & [-20pt] D_1 \arrow[dash]{r}& [-20pt] D_2
\end{tikzcd}
\]
where critically, there is an extra divisor $D_2$ corresponding to an additional tensor multiplet. On the other hand, the case without vector structure can be realized in the frozen phase of F-theory with the following configuration:
\[
\begin{tikzcd}
\text{Frozen:}\quad~\left [\fsp_{8}\right ] &[-35pt]\overset{\mathfrak{su}_{24}}{0}&[-10pt] [\frac{1}{2} F] \arrow[dash]{r} & [-20pt] 2D_1 
\end{tikzcd}
\]

We note that the corresponding geometry is obtained by simply contracting the divisor $D_2$. 

These two cases lead to a natural generalization for a description of small $\textit{Spin}(32)/\mathbbm{Z}_2$-instantons on arbitrary $A_{2n-1}$-singularities. Indeed, the case of an $A_3$-singularity with and without vector structure corresponding to the following F-theory realizations in the unfrozen and the frozen phases of F-theory respectively:
\[
\begin{tikzcd}[row sep=0mm]
&[-5pt]\text{Unfrozen:}&\left [\fso_{32}\right ] &[-35pt]\overset{\mathfrak{sp}_{24}}{1}&[-35pt]\overset{\mathfrak{su}_{40}}{2}&[-35pt]\overset{\mathfrak{sp}_{16}}{1}&[-10pt] [ F] \arrow[dash]{r} & [-20pt] D_1\arrow[dash]{r}  & [-20pt] D_2 \arrow[dash]{r} & [-20pt] D_3\\[0pt]
&[-5pt]\text{Frozen:}~~\;&\left [\fsp_{8}\right ] &[-35pt]\overset{\mathfrak{su}_{24}}{1}&[-35pt]\overset{\mathfrak{su}_{16}}{2}&&[-10pt] [ F] \arrow[dash]{r} & [-20pt] (2D_1+D_2)\arrow[dash]{r}  & [-20pt] D_2
\end{tikzcd}
\]
In complete analogy with the $A_1$-case, the frozen geometry is obtained by simply contracting the last divisor $D_3$.

For the general situation of $24$ small $SO(32)$ instantons on an $A_{2n-1}$-singularity, we claim that the respective F-theory geometries are realized by the following chains:
\[
\begin{tikzcd}
\text{Unfrozen:}\quad\left [\fso_{32}\right ] &[-35pt]\overset{\mathfrak{sp}_{24}}{1}&[-35pt]\overset{\mathfrak{su}_{40}}{2}&[-35pt]\ldots & [-35pt] \overset{\mathfrak{su}_{64 - 8n}}{2}& [-35pt] \overset{\mathfrak{sp}_{28-4n}}{1}&[-10pt] [ F] \arrow[dash]{r} & [-20pt] D_1\arrow[dash]{r}  & [-20pt] D_2 \arrow[dash]{r}& [-20pt] \ldots \arrow[dash]{r}& [-20pt] D_{n-1}\arrow[dash]{r} &[-20pt] D_n\,,
\end{tikzcd}
\]
\[
\begin{tikzcd}
\text{Frozen:}\quad\left [\fsp_{8}\right ] &[-35pt]\overset{\mathfrak{su}_{24}}{1}&[-35pt]\ldots & [-35pt] \overset{\mathfrak{su}_{40-8n}}{1}&[-10pt] [ \frac{1}{2}F] \arrow[dash]{r} & [-20pt] (2D_1 + D_2 + \ldots + D_n)\arrow[dash]{r} & [-20pt] D_2 \arrow[dash]{r}& [-20pt] \ldots\arrow[dash]{r} &[-20pt] D_n\,,
\end{tikzcd}
\]
where the frozen F-theory geometry is again obtained by contracting the divisor $D_n$ in the base.

\subsubsection[\texorpdfstring{$D_n$ and $E_m$}{Dn and Em} singularities]{\texorpdfstring{$\boldsymbol{D_n}$ and $\boldsymbol{E_m}$}{Dn and Em} singularities}
Our constructions for the frozen phase of F-theory describing small $SO(32)$-instantons on $A_{2n-1}$ singularities without vector structure can be readily generalized to the cases of $D$ and $E$ singularities. In this section, we will briefly describe the results for $D_n$ and $E_6$ singularities.

As in the previous section, we will describe explicitly only the cases with the maximal number of $24$ small instantons probing $D$- and $E$-type singularities. We begin with the case of $D_4$ singularities, where both cases with and without vector structure were derived in Section~\ref{sec:Dnsing}. The quivers describing the gauge symmetries and matter content were given, respectively, by the following:
\[
\begin{tikzcd}[row sep=0pt]
&[-35pt]&[-35pt]&[-35pt] \overset{\mathfrak{sp}_{16}}{1} & [-35pt]&[-10pt] & [-20pt]& E_1\arrow[dash]{d} \\[-4pt]
\text{Unfrozen:}\quad&\left [\fso_{32}\right ] &[-35pt]\overset{\mathfrak{sp}_{24}}{1}&[-35pt]\overset{\mathfrak{so}_{80}}{4}&[-35pt] \overset{\mathfrak{sp}_{16}}{1}&[-10pt] [ F] \arrow[dash]{r} & [-20pt] D_1\arrow[dash]{r}  & [-20pt] D_2 \arrow[dash]{r}&[-20pt] D_3 \\[-4pt]
&[-35pt]&[-35pt]&[-35pt] \overset{\mathfrak{sp}_{16}}{1} & [-10pt] & [-35pt] & [-20pt] & E_2 \arrow[dash]{u}\\
\text{Frozen:}~~\;\quad&\left [\fsp_{8}\right ] &[-35pt]\overset{\mathfrak{su}_{24}}{2}&[-35pt]\overset{\mathfrak{sp}_{16}}{1}&[-35pt] \overset{\mathfrak{su}_{16}}{2}& \left[ \frac{1}{2}F\right] \arrow[dash]{r} & [-20pt] (2D_1+D_2)\arrow[dash]{r}  & [-20pt] D_3 \arrow[dash]{r}&[-20pt] D_2
\end{tikzcd}
\]
The diagrams show the relevant gauge algebras, their intersection properties, and the intersection numbers of their support, which uniquely determines the matter content by anomaly cancellation.

The first diagram above can be constructed in an ordinary F-theory compactification on an elliptic fibration $X$ over a blowup, $B$, of $\mathbbm{F}_4$, with $\mathfrak{so}(32)$ supported on the proper transform of the $(-4)$-curve. The second diagram can be obtained from a frozen F-theory compactification on $X$ as follows. First, one blows down the $(-1)$-curves labeled as $E_i$ in the first diagram in the base $B$. Second, one constructs a frozen F-theory embedding with the gauge divisors specified as in the second diagram above, and one can verify that these are indeed consistent with the rules described in Section~\ref{sec:f}.

Our analysis for the case of $D_5$-singularities proceeds almost identically; the relevant gauge algebras and intersection patterns in the cases with and without vector are given by the following:
\[
\begin{tikzcd}[row sep=0pt]
&[-35pt]&[-35pt]\overset{\mathfrak{sp}_{16}}{1}&[-35pt]&[-35pt] & [-10pt]&[-20pt] & [-20pt] E_1\arrow[dash]{d} \\[-4pt]
\text{Unfrozen:}\quad&\left [\fso_{32}\right ] &[-35pt]\overset{\mathfrak{sp}_{24}}{1}&[-35pt]\overset{\mathfrak{so}_{80}}{4}&[-35pt] \overset{\mathfrak{sp}_{32}}{1}~ \overset{\mathfrak{su}_{32}}{2}&[-10pt] [ F] \arrow[dash]{r} & [-20pt] D_1\arrow[dash]{r}  & [-20pt] D_2 \arrow[dash]{r}&[-20pt] E_2  \arrow[dash]{r}& [-20pt] D_3\\
\end{tikzcd}
\]
\[
\begin{tikzcd}[row sep=0pt]
~~\text{Frozen:}&\left [\fsp_{8}\right ] &[-35pt]\overset{\mathfrak{su}_{24}}{2}&[-35pt]\overset{\mathfrak{su}_{32}}{1}&[-35pt] \overset{\mathfrak{su}_{16}}{2}&[-10pt] [ \frac{1}{2}F] \arrow[dash]{r} & [-20pt] (2D_1+D_2)\arrow[dash]{r}  & [-20pt] D_3 \arrow[dash]{r}&[-20pt] D_2 
\end{tikzcd}
\]
As in the above, the frozen configuration is realized by shrinking the unlabeled divisors in the first diagram, and then choosing the displayed assignment of gauge divisors.

Finally, we give the general assignment for arbitrary $D_n$ singularities, dividing into the cases where $n$ is even or odd. In the even case, we find the following for 
$D_{2n+4}$:\\[12pt]
\text{Unfrozen:}
\[
\begin{tikzcd}
&[-35pt]&[-35pt]\overset{\mathfrak{sp}_{16}}{1}&[-35pt]&[-35pt]&[-35pt]&[-35pt]\overset{\mathfrak{sp}_{16-4n}}{1}\\[-25pt]
\left [\fso_{32}\right ] &[-35pt]\overset{\mathfrak{sp}_{24}}{1}&[-35pt]\overset{\mathfrak{so}_{80}}{4}&[-35pt]\overset{\mathfrak{sp}_{32}}{1}&[-35pt]\ldots &[-35pt]\overset{\mathfrak{sp}_{40-8n}}{1}&[-35pt]\overset{\mathfrak{so}_{80-16n}}{4}&[-35pt]\overset{\mathfrak{sp}_{16-4n}}{2}
\end{tikzcd}
\begin{tikzcd}[column sep = .3cm, row sep = .3cm]
 &&E_1\arrow[dash]{d}&&& &E_2\arrow[dash]{d}&\\
\left[F\right]\arrow[dash]{r} &D_1\arrow[dash]{r} &D_2\arrow[dash]{r} &E_3\arrow[dash]{r}&\ldots \arrow[dash]{r}&E_k\arrow[dash]{r}&D_{n-1}\arrow[dash]{r} &D_n
\end{tikzcd}
\]
\text{Frozen:}
\[
\begin{tikzcd}
\left[\mathfrak{sp}_8\right] & [-35pt]\overset{\mathfrak{su}_{24}}{2} & [-35pt] & [-35pt] & [-35pt] & [-35pt] & [-35pt] \\ [-20pt]
 & [-35pt] \overset{\mathfrak{su}_{32}}{2} & [-35pt] \overset{\mathfrak{su}_{24}}{1} & [-35pt] \ldots & [-35pt]  \overset{\mathfrak{su}_{40-8n}}{2} & [-35pt]  \overset{\mathfrak{sp}_{16-4n}}{1}\\ [-20pt]
 & [-35pt]  \overset{\mathfrak{sp}_{16}}{2}
\end{tikzcd}
\hspace*{1cm}
\begin{tikzcd}[column sep = .3cm, row sep = .3cm]
\left[\frac{1}{2}F\right]\arrow[dash]{r} &2D_1+ D_2\arrow[dash]{d} &&& \\
& D_3 \arrow[dash]{d}\arrow[dash]{r}& D_4\arrow[dash]{r}& \ldots \arrow[dash]{r}& D_{n-1} \arrow[dash]{r} & D_n \\
& D_2
\end{tikzcd}
\]
Similarly, in the odd cases, we have the following diagrams for $D_{2n+5}$:\\[12pt]
\[
\adjustbox{scale=.93}{%
\begin{tikzcd}
&[-35pt]&[-35pt]\overset{\mathfrak{sp}_{16}}{1}&[-35pt]&[-35pt]&[-35pt]\\[-25pt]
\text{Unfrozen:}\quad&\left [\fso_{32}\right ] &[-35pt]\overset{\mathfrak{sp}_{24}}{1}&[-35pt]\overset{\mathfrak{so}_{80}}{4}&[-35pt]\ldots &[-35pt]\overset{\mathfrak{so}_{80-16n}}{4}&[-35pt]\overset{\mathfrak{sp}_{16-4n}}{1}&[-35pt]\overset{\mathfrak{su}_{16-4n}}{2}
\end{tikzcd}
\begin{tikzcd}[column sep = .3cm, row sep = .3cm]
 &&E_1\arrow[dash]{d}&&& &&\\
\left[F\right]\arrow[dash]{r} &D_1\arrow[dash]{r} &D_2\arrow[dash]{r}&\ldots \arrow[dash]{r}&D_{n-1}\arrow[dash]{r}&E_k\arrow[dash]{r} &D_n\\
\end{tikzcd}
}%
\]
\[
\adjustbox{scale=.93}{%
\begin{tikzcd}
\text{Frozen:}\quad&\left[ \fsp_8 \right]&[-35pt]\overset{\mathfrak{su}_{24}}{2}&[-35pt]&[-35pt]&[-35pt]\\[-25pt]
&& \overset{\mathfrak{su}_{32}}{2} & [-35pt] \overset{\mathfrak{su}_{24}}{1} & [-35pt]\ldots & [-35pt] \overset{\mathfrak{su}_{40-8n}}{2} & [-35pt]\overset{\mathfrak{su}_{16-4n}}{1} \\[-20pt]
&& \overset{\mathfrak{su}_{16}}{2}
\end{tikzcd}
\hspace*{.5cm}
\begin{tikzcd}[column sep = .3cm, row sep = .3cm]
\left[\frac{1}{2} F\right]\arrow[dash]{r} &2D_1+ D_2 \arrow[dash]{d} \\
 & D_3\arrow[dash]{r} \arrow[dash]{d}  & D_4 \arrow[dash]{r} & \ldots \arrow[dash]{r}  & D_{n-1} \arrow[dash]{r} & D_n \\
 & D_2
\end{tikzcd}
}%
\]

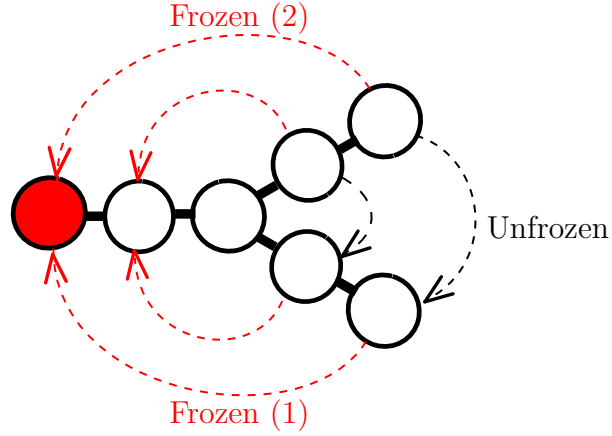
\begin{figure}[t!]
\begin{center}
\begin{picture}(0,170)
\put(-50,20){\includegraphics{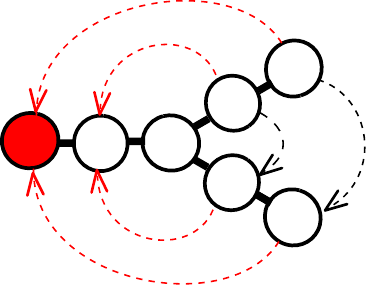}}
\put(10,10){\textcolor{red}{Frozen $(1)$}}
\put(10,160){\textcolor{red}{Frozen $(2)$}}
\put(130,80){Unfrozen}
\end{picture}
\caption{The $\fe_6^{(1)}$ quiver base shape of $\fso_{32}$ instantons probing the same singularity, with affine node in red. There are three choices of $\mathbbm{Z}_2$ foldings. Black arrows highlight the $\mathbbm{Z}_2$ folding in the case with vector structure, and the two red arrows show the folding when vector structure is switched off.}
\label{fig:E61folding}
\end{center}
\end{figure}

We now consider $E_6$-singularities. Similar to the $D_{2n+3}$ and $A_{2n}$ cases, the quiver in the unfrozen phase is not that of affine $E_6$, but is folded by its $\mathbbm{Z}_2$ automorphism to the quiver of $F_4$~\cite{Blum:1997mm} (see Table~\ref{fig:LSTQuivers}). The gauge group chain reads
\[
\begin{tikzcd}
\text{Unfrozen:}~\quad\left [\fso_{32}\right ] &[-35pt]\overset{\mathfrak{sp}_{24}}{1}&[-35pt]\overset{\mathfrak{so}_{80}}{4}&[-35pt] \overset{\mathfrak{sp}_{48}}{1}&[-35pt] \overset{\mathfrak{su}_{64}}{2}&[-35pt] \overset{\mathfrak{su}_{32}}{2}
\end{tikzcd}
\hspace*{1cm}
\begin{tikzcd}[column sep = .3cm, row sep = .3cm]
\left[ F\right] \arrow[dash]{r} & D_1\arrow[dash]{r}  & D_2 \arrow[dash]{r}&D_3  \arrow[dash]{r}&D_4\arrow[dash]{r}&D_5
\end{tikzcd}
\]
In Section~\ref{sec:Ansing}, Table~\ref{fig:LSTQuivers} we observed that switching off vector structure in $G$ orbifolds of $\textit{Spin}(32)/\mathbbm{Z}_2$ instantons amounts to a $\mathbbm{Z}_2$-folding of the affine diagram. Due to the triality of affine $E_6$, this folding also leads to a quiver with the shape of affine $F_4$, given by the frozen divisor combination\\[12pt]
\text{Frozen (I):}
\[
\adjustbox{scale=.9}{%
\begin{tikzcd}
\left [\fsp_{8}\right ] &[-35pt]\overset{\mathfrak{su}_{24}}{2}&[-35pt]\overset{\mathfrak{su}_{32}}{2}&[-35pt] \overset{\mathfrak{sp}_{20}}{1}&[-35pt] \overset{\mathfrak{so}_{32}}{4}&[-35pt] \overset{\mathfrak{sp}_{4}}{1}
\end{tikzcd}
}\hspace*{1cm}
\adjustbox{scale=.9}{%
\begin{tikzcd}[column sep = .3cm, row sep = .3cm]
\left[ \frac{1}{2}F\right] \arrow[dash]{r} &2D_1 + D_2 + 2 D_3 + D_4\arrow[dash]{r}  & D_4 + D_5 \arrow[dash]{r}& D_3  \arrow[dash]{r}&  D_2\arrow[dash]{r}& D_3 + D_4
\end{tikzcd}
}
\]
Moreover, there exists a third possibility to fold the affine $E_6$ to $F_4$ (see Figure~\ref{fig:E61folding}), and thus we expect a second frozen F-theory divisor assignment. This does indeed exist,
\[
\begin{tikzcd}[column sep = .3cm, row sep = .3cm]
\text{Frozen (II):}\quad\left[ \frac{1}{2}F\right] \arrow[dash]{r} &2D_1 + D_2 + 2 D_3 + D_4+ D_5\arrow[dash]{r}  & D_4  \arrow[dash]{r}& D_3  \arrow[dash]{r}&  D_2\arrow[dash]{r}& D_3 + D_4 + D_5
\end{tikzcd}
\]
while the gauge group chain is exactly the same as before. From the 6D SUGRA perspective both quivers look exactly the same, while geometrically they arise from two different frozen brane assignments. It would be interesting to investigate whether the theories differ by more than just the inequivalent two frozen brane assignments.

Finally, we consider $E_7$ singularities, for which the divisor assignments read\\[12pt]
\text{Unfrozen:}
\[
\begin{tikzcd}
&[-35pt]&[-35pt]&[-35pt]&[-35pt]\overset{\mathfrak{sp}_{28}}{1}&[-35pt]&[-35pt]&[-35pt]\\[-25pt]
\left [\fso_{32}\right ] &[-35pt]\overset{\mathfrak{sp}_{24}}{1}&[-35pt]\overset{\mathfrak{so}_{80}}{4}&[-35pt] \overset{\mathfrak{sp}_{48}}{1}&[-35pt] \overset{\mathfrak{so}_{128}}{4}&[-35pt] \overset{\mathfrak{sp}_{44}}{1}&[-35pt] \overset{\mathfrak{so}_{64}}{4}&[-35pt] \overset{\mathfrak{sp}_{12}}{1}
\end{tikzcd}
\hspace*{1cm}
\begin{tikzcd}[column sep = .3cm, row sep = .3cm]
 &&&&E_1\arrow[dash]{d}&&\\
\left[F\right]\arrow[dash]{r} &D_1\arrow[dash]{r} &D_2\arrow[dash]{r} &D_3\arrow[dash]{r} \arrow[dash]{r}&D_{4}\arrow[dash]{r} &E_2\arrow[dash]{r} &D_5\arrow[dash]{r} &E_3
\end{tikzcd}
\]
\text{Frozen:}
\[
\adjustbox{scale=1}{%
\begin{tikzcd}
\left [\fsp_{8}\right ] &[-35pt]\overset{\mathfrak{su}_{24}}{2}&[-35pt]\overset{\mathfrak{su}_{32}}{2}&[-35pt] \overset{\mathfrak{su}_{40}}{2}&[-35pt] \overset{\mathfrak{sp}_{24}}{1}&[-35pt] \overset{\mathfrak{so}_{32}}{4}
\end{tikzcd}
}
\hspace*{1cm}
\begin{tikzcd}[column sep = .3cm, row sep = .3cm]
\left[F\right]\arrow[dash]{r} &2D_1 + D_2 + 2 D_3 + D_4\arrow[dash]{r} &D_5\arrow[dash]{r} &D_4\arrow[dash]{r} \arrow[dash]{r}&D_{3}\arrow[dash]{r} &D_2
\end{tikzcd}
\]

\subsection[Higgsings of \texorpdfstring{$SO(32)$}{SO(32)}]{Higgsings of \texorpdfstring{$\boldsymbol{SO(32)}$}{SO(32)}}
In the previous two sections, we carefully analyzed a variety of singular limits and phase transitions for the $SO(32)$ heterotic string without vector structure dual to frozen F-theory compactifications on $\mathbbm{F}_4$ with an $\mathfrak{sp}_{8}$ gauge symmetry. Such configurations necessarily contain highly non-perturbative dynamics, such as small instantons probing ADE singularities, but the full details of the physics is inaccessible purely from the perturbative heterotic string. Our goal in this section is to present some simple examples of Higgsings of the frozen F-theory compactification described above, while a full exploration of the duality to the perturbative heterotic string without vector structure will be left for future work.

Motivated by examples of perturbative heterotic compactifications with $SO(28) \times SU(2)$ gauge symmetry, we will consider the Higgsings of the $SO(32)$ gauge symmetry to $SO(28)$, and the natural freezings of this configuration. We begin by considering the following tuning of Weierstrass coefficients on $\mathbbm{F}_4$:
\begin{align*}
f &= \frac{1}{3} e^2 (-p_2^2 + 3 e^5 h p_4)\,,\quad g = \frac{1}{27} e^3  p_2 (-2 p_2^2 + 9 e^5 h p_4)\,,\quad \Delta = e^{16} h^2 p_4^2 (-p_2^2 + 4 e^5 h p_4)\,,
\end{align*}
where $p_2 \in \vert 3e + 12f \vert$, and $p_4 \in \vert 20f \vert $, where $\vert D \vert$ denotes the linear system associated with a divisor $D$. Taking $-p_2^2 + 4e^5 h p_4 = 0$ to be the residual locus $\widetilde{\Delta}$, we have that $\widetilde{\Delta} \in \vert 6e + 24f \vert$ and the intersection relations
\[
\widetilde{\Delta} \cdot e = 0, \quad \widetilde{\Delta} \cdot f = 6, \quad \widetilde{\Delta} \cdot h = 24
\]

In the unfrozen setting, we obtain the following gauge algebras and matter assignments
\begin{itemize}
\item
$\mathfrak{so}_{28}$ on $e$, $\mathfrak{su}_2$ on $h$, $\mathfrak{sp}_{20}$ on $p_4$
\item
$1$ half-bifundamental of $\mathbf{28} \otimes \mathbf{40}$
\item
$1$ bifundamental of $\mathbf{40} \otimes \mathbf{2}$
\item
$1$ anti-symmetric of $\mathfrak{sp}_{20}$
\item
$1$ tensor multiplet, $25$ neutral hypermultiplets
\end{itemize}

We now flip to a frozen F-theory configuration, where $e$ supports a frozen divisor $\widehat{I}_{10}^*$. We claim the following matter assignments:
\begin{itemize}
\item
$\mathfrak{sp}_6$ on $\frac{1}{2} e$, $\mathfrak{su}_2$ on $h$, $\mathfrak{sp}_{10}$ on $2f$
\item
$1$ bifundamental of $\mathbf{12} \otimes \mathbf{20}$
\item
$2$ bifundamentals of $\mathbf{2} \otimes \mathbf{20}$
\item
$1$ antisymmetric of $\mathfrak{sp}_{10}$
\item
$1$ tensor multiplet, $25$ neutral hypermultiplets
\end{itemize}

We note that the number of neutral hypermultiplets before and after freezing are precisely the same, as we would expect naively from the geometry, consistent with the discussions of Section~\ref{sec:summary}.

\section{Fusing SUGRA Theories}
\label{sec:frozengravity}
The 6D instanton theories constructed in Section~\ref{sec:heteroticwovector} are non-gravitational theories that flow to little string theories in the UV. When discussing their frozen F-theory duals in the previous section, we have considered them as subsectors of full 6D SUGRA theories. In the following, we generalize this construction of \textit{fusion} of two 6D $\textit{Spin}(32)/\mathbbm{Z}_2$-instanton theories along their flavor symmetries with and without vector structure to obtain a broader class of $6$d frozen F-theory vacua.

\subsection{Fusion of Little Strings}
The process of fusion follows the very same concept from 6D SCFTs~\cite{Heckman:2018pqx} and refers to coupling two different SCFTs $\mathcal{T}_1$ and $\mathcal{T}_2$, along a common flavor group factor $[\fg_F]$
\begin{align}
 \mathcal{T}_3 =
\mathcal{T}_1 
\textcolor{blue}{[ \mathfrak{g}_F ]} 
\mathrel{\stackon[-1pt]{{-}\mkern-5mu{-}\mkern-5mu{-}\mkern-5mu{-}}{\textcolor{blue}{ \mathfrak{g}_F }}}
\textcolor{blue}{[  \mathfrak{g}_F ]} \,   
\mathcal{T}_2 \, ,
\end{align}
to obtain a new SCFT $\mathcal{T}_3$. Depending
on the details of the fusion process, the resulting theory may change substantially its UV behavior. In particular, it may not flow to an SCFT anymore but to a 6D Little String Theory \cite{Bhardwaj:2015oru,DelZotto:2022xrh} or a SUGRA theory instead. 

In the following, we describe the fusion process of two non-compact Little String Theories from the class $\mathcal{K}^{UF}_M(G)$ with vector structure, or from $\mathcal{K}_M^F(G)$ without vector structure, as summarized in Table~\ref{fig:LSTQuivers}. The quivers are fused by gauging a common flavor algebra $\fg_F$, which introduces the respective vector multiplets and a tensor multiplet. The resulting new quiver $\mathcal{G}(G_1,G_2,M)^{UF/F}$ that describes the SUGRA theory is
\begin{align}
 \mathcal{G}^{UF/F}(G_1,G_2,M) =
\mathcal{K}^{UF/F}_M(G_1)
\textcolor{blue}{[ \mathfrak{g}_F ]} 
\mathrel{\stackon[-1pt]{{-}\mkern-5mu{-}\mkern-5mu{-}\mkern-5mu{-}}{\textcolor{blue}{ \mathfrak{g}_F }}}
\textcolor{blue}{[  \mathfrak{g}_F ]} \,   
\mathcal{K}^{UF/F}_{N}(G_2) \, ,
\end{align}
where the fusion flavor algebra is highlighted in blue. 

The fusion comes with a set of conditions needed to obtain a consistent 6D SUGRA theory. In particular, all global symmetries need to be either trivialized or gauged. Little string theories can have the following two types of global symmetries:
\begin{itemize}
\item Flavor symmetries of type $\fso_{32}$ or $\fsp_8$ in the case of vector structure or no vector structure.
\item A global $\mathfrak{u}_1$ 1-form symmetry acting on tensor multiplets.\footnote{One may also worry about 2-form defect group factors \cite{DelZotto:2015isa} that may be gauged as well \cite{Braun:2021sex}. However, it can be readily checked that none of the heterotic little string theories admit a non-trivial 2-form symmetry group~\cite{Bhardwaj:2020phs}.}
\end{itemize}
While fusion gauges the $\fso_{32}/\fsp_8$ flavor symmetries, the $\mathfrak{u}_1$ 1-form symmetry is still left. Higher global symmetries will be absent in 6D strings coupled to tensor multiplets if the intersection form $\Omega$ of the F-Theory base is integral and self-dual~\cite{Seiberg:2011dr}. Since each heterotic LST building block is birational to a $0$ curve, we find that the minimal blow-down base must be a Hirzebruch surface $\mathbbm{F}_n$, where $n$ is the negative self-intersection of the fusion curve. In the unfrozen
case (and the full $\fso_{32}$ algebra unbroken), this is an $\overset{\fso_{32}}{n}$. Anomalies fix $n=4$ and demand $24$ hypermultiplets in the vector representation. This fixes the quivers to have the shapes
\begin{align}
\ldots \overset{\fsp_{24-M}}{1} \, \,\textcolor{blue}{\overset{\fso_{32}}{4}}  \, \, \overset{\fsp_{M}}{1} \ldots  \,.
\end{align}
Note that the sum of the instanton contributions is 24 as expected for a heterotic string. The integer $M$ labels the relative number of instantons on either side of the $\fso_{32}$ and gives a bound on the number of inequivalent models.
 
Fusion works similarly for the frozen models. Instead of the versions with vector structure, we freeze the quivers $\mathcal{K}^{UF}_M(G) \rightarrow \mathcal{K}^F_M(G)$ and pursue in the same manner. Note that the frozen 7-brane $F$ always lies on the class of the fusion node. Hence, when freezing the class to $\frac12F$, the $\fso_{32}$ is reduced to $\fsp_{8}$ and the BPS string charge given by $F^2=4$ reduced to $(\frac12 F)^2=1$ as expected. Anomaly freedom requires again $24$ instantons, such that the resulting quiver shapes are
\begin{align}
\ldots \overset{\fsu_{24-M}}{n_1} \, \, \textcolor{blue}{\overset{\fsp_8}{1}} \, \, \overset{\fsu_{M}}{n_2} \ldots  \, .
\end{align} 

The respective 6D SUGRA theories can therefore be labeled by two types of orbifold singularities $G_1, G_2$, their relative number of five-branes $M$ and a label that indicates whether they are frozen,
\begin{align}
\mathcal{G}^{UF}(G_1,G_2,M) \, , \qquad \text{ or } \qquad \mathcal{G}^F(G_1,G_2,M) \, .
\end{align}
It is important that freezing keeps the structure of the heterotic LSTs, i.e., that they are  birational to $0$ curves.\footnote{On the level of the little string theory, this is equivalent that freezing keeps the 2-group structure constant $\kappa_P$ invariant \cite{Cordova:2020tij}.} Hence, fusion is independent of freezing. Since the fusion node is always in the class $F$ of the $O7^+$, we cannot fuse a frozen and an unfrozen theory. This is also clear from requiring that the fusion nodes have the same flavor symmetry, which is not the case for a frozen and an unfrozen model. 

The process of fusion and freezing can be summarized in the commuting diagram 
\begin{align}
\begin{array}{ccccc}
\mathcal{K}^{UF}(G_1)_M & \xrightarrow{\text{Fusion}} & \mathcal{G}^{UF}(G_1,G_2,M) &\xleftarrow{\text{Fusion}}& \mathcal{K}(G_2)^{UF}_{24-M}  \\
\Bigg\downarrow \text{\scriptsize freeze}& & \Bigg\downarrow && \text{\scriptsize freeze} \Bigg\downarrow \\  
\mathcal{K}^{F}(G_1)_M & \xrightarrow{\text{Fusion}} & \mathcal{G}^{F}(G_1,G_2,M) &\xleftarrow{\text{Fusion}}& \mathcal{K}(G_2)^{F}_{24-M}  \\
\end{array} \,
\end{align} 

Fusing two instanton theories into a gravity theory readily implies a couple of interesting facts. First, we will show that the number of (toric) complex structure moduli is constant, as can be derived from the field theory properties of $\fso_{32}$ instanton theories. The dimension of the moduli space of $M$ $\fso_{32}$ instantons ~\cite{Blum:1997mm} on a $G$ singularity is the same as its Higgs branch dimension and given by 
\begin{align}
\label{eq:hbranch}
\text{dim}_H ( \mathcal{K}_M(G_1) ) = c(G) + 30 M \, ,
\end{align} 
with $c(G)$ some $G$-dependent constant. This Higgs branch dimension, on the other hand, is basically the gravitational anomaly coefficient
\begin{align}
\text{dim}_H = H-V +29 T \, ,
\end{align}
which when coupled to gravity, must equal
\begin{align}
H_{\mathbf{1}} + \text{dim}_H = 273  \, ,
\end{align}
with $H_{\mathbf{1}}$ the number of uncharged hypermultiplets. The latter coincide with the number of complex structure moduli of the respective geometry up to the universal hypermultiplet. When fusing two instanton theories $\mathcal{K}_M (G_1) $ and $\mathcal{K}_N (G_2) $ to a SUGRA theory, their respective Higgs branch dimensions are added, and one introduces $496$ vectors of $\fso_{32}$ and its tensor multiplet contribution. The instanton numbers $N,M$ of the fusion constituents are constrained to lead to the $24$ heterotic instantons, which requires fixing $N= c_2 -M$ with $c_2$ some constant. The number $M$ therefore only labels the \textit{relative} instanton number $M$ between the two fused theories. Using the gravitation anomaly, we then have
\begin{align}
\mathbf{H_{1}}=273-\text{dim}_H ( \mathcal{K}_M(G_1))- \text{dim}_H (  \mathcal{K}_{c_2-M} (G_2))-29 + \text{dim}(\fso_{32}) \, .
\end{align} 
When inserting this into \eqref{eq:hbranch}, we find that $M$ cancels and we are left with a constant number of neutral hypermultiplets. Another way of seeing this is to note that the number of neutral hypermultiplets is related to the number of (polynomial) complex structure deformations of the respective threefold geometries. However, since freezing is just a \textit{re-interpretation} of the same Weierstrass model, its polynomial deformation should be the same, as discussed Section~\ref{sec:summary}, and the number of neutral fields is invariant. 

Notably, there is an exception when the singularity of either instanton theory is chosen to be trivial, $\fg=\emptyset$. In such cases we will find all deformations to be of non-polynomial origin, i.e., not visible in the hypersurface equation.

\subsection{Examples} 
In the following, we give concrete examples for the fusion algorithm outlined above. We start by constructing an unfrozen F-theory SUGRA model obtained by fusing two unfrozen heterotic instanton theories from Table~\ref{fig:LSTQuivers}, which we will then freeze in the second step. 

\paragraph{$\boldsymbol{\mathcal{G}_M (\emptyset, D_4)}$ models.}
We start by picking a $D_4$-type singularity and fuse it with an unorbifolded instanton theory, which yields the quiver
\begin{align}
\label{eq:-d4UFQuiver}
\overset{\fsp_{16-M}}{0}  \, \, \textcolor{blue}{ \overset{\fso_{32}}{4} }\, \, \overset{\fsp_{8+M}}{1} \, \,  \overset{ \overset{\fsp_M}{ 1}}{ \underset{\overset{\fsp_M}{1}}{\overset{\fso_{16+4M}}{4}}}  \, \, \overset{\fsp_{M}}{1} \, ,
\end{align} 
valid for $M=0 \ldots 16$. The above model can be  constructed via a hypersurface in a 4D reflexive polytope as spelled out in Appendix~\ref{app:ToricData}. From those vertices we can compute the the toric hypersurface as well as the Hodge numbers,
\begin{align}
(h^{1,1},h^{2,1})(X_{M,D_4})=(55+5M, 31-M) \, ,
\end{align} 
which matches the rank of the above gauge algebra and SUGRA anomaly cancellation. Note that for the number of neutral hypermultiplets 
$H_{n}=h^{2,1}+1$ in 6D, we also need to take into account that the anti-symmetric representation of $\fsp_n$ of dimension $
\mathbf{n (2 n - 1) - 1}$ 
does not have full \textit{charge dimension}. The number of charged hypermultiplets in this irrep is
\begin{align}
\text{dim}(\mathbf{n(2 n - 1) - 1}) - \text{dim}(R_0) = 2 (n-1) n \, .
\end{align}
Here $\text{dim}(R_0)=n-1$ denotes the number of weights of the antisymmetric representation that have trivial charge under all Cartan generators. In the geometry of the respective threefolds, these are already accounted for in $h^{2,1}$. The corresponding complex structure deformations are non-torically realized, meaning they do not appear as monomials in the hypersurface equation.\footnote{More details of an F/M-theory interpretation of these deformations are given in \cite{Anderson:2023wkr}.} The polynomial complex structure coefficients, on the other hand, are in fact constant for any value of $M$ and given by\footnote{The values differ slightly for $M=15$ and $M=16$, in which case no antisymmetric irreps are present.}
\begin{align}
h^{2,1}_\text{toric}(X_{M,D_4})= 16 \, .
\end{align}  
Having established the unfrozen model, we can simply freeze it by exchanging the heterotic instanton theories on both sides by their versions without vector structure. The respective frozen model is given as 
 \begin{align} 
\overset{\fsp_{16-M}}{0}  \, \, \textcolor{blue}{ \overset{\fso_{32}}{4} }\, \, \overset{\fsp_{8-M}}{1} \, \,  \overset{ \overset{\fsp_M}{ 1}}{ \underset{\overset{\fsp_M}{1}}{\overset{\fso_{16+4M}}{4}}}  \, \, \overset{\fsp_{M}}{1} \,  \qquad  \xrightarrow{\text{freeze}} \quad  \overset{\fsp_{8-M/2}}{0}  \, \, \textcolor{blue}{ \overset{\fsp_{8}}{1}}  \, \, \overset{\fsu_{8+M}}{2} \, \,  \overset{\fsp_{M}}{1 }  \, \, \overset{\fsu_{M}}{2} \, ,
\end{align}
which requires the relative number of instantons $M$ to be even. It is easy to verify that the number of neutral hypermultiplets in the above model remains unchanged. 

\paragraph{$\boldsymbol{\mathcal{G}_M (D_4, D_4)}$ models.}
Moving on to more complicated examples, we next fuse two $D_4$-type theories, which results in the SUGRA quiver 
\begin{align}
\label{eq:D4D4UFquiver}
  \overset{\fsp_{8-M}}{1} \, \,  \overset{ \overset{\fsp_{8-M}}{ 1}}{ \underset{\overset{\fsp_{8-M}}{1}}{\overset{\fso_{48-4M}}{4}}}  \, \, \overset{\fsp_{16-M}}{1}  
  \, \, \textcolor{blue}{ \overset{\fso_{32}}{4} }\, \, \overset{\fsp_{8+M}}{1} \, \,  \overset{ \overset{\fsp_M}{ 1}}{ \underset{\overset{\fsp_M}{1}}{\overset{\fso_{16+4M}}{4}}}  \, \, \overset{\fsp_{M}}{1} \, ,
\end{align}
for $M=0 \ldots 8$. The model can be constructed torically, from which we compute the Hodge numbers
\begin{align}
(h^{1,1},h^{2,1})(X_M)=(107,11) \, ,
\end{align}
consistent with the 6D SUGRA anomalies. This time all complex structure moduli are polynomial and constant for any $M$ since this model does not admit any anti-symmetric representations. Moreover, the number of K\"ahler moduli is also constant since the ranks of the gauge algebras on the left and right side of the $\fso_{32}$ exactly cancel out. This always happens when there are two identical singularities, i.e., for $\mathcal{G}_M (G, G)$ theories. Freezing then results in 
\begin{align}
\label{eq:D4D4Fquiver}
  \overset{\fsp_{8-M}}{1} \, \,  \overset{ \overset{\fsp_{8-M}}{ 1}}{ \underset{\overset{\fsp_{8-M}}{1}}{\overset{\fso_{48-4M}}{4}}}  \, \, \overset{\fsp_{16-M}}{1}  
  \, \, \textcolor{blue}{ \overset{\fso_{32}}{4} }\, \, \overset{\fsp_{8+M}}{1} \, \,  \overset{ \overset{\fsp_M}{ 1}}{ \underset{\overset{\fsp_M}{1}}{\overset{\fso_{16+4M}}{4}}}  \, \, \overset{\fsp_{M}}{1}  \qquad \xrightarrow{\text{freeze}} \quad   
  \overset{\fsu_{8-M}}{2} \, \,  \overset{\fsp_{8-M}}{1 }  \, \, \overset{\fsu_{16-M}}{2}
 \, \, \textcolor{blue}{ \overset{\fsp_{8}}{1}}  \, \, \overset{\fsu_{8+M}}{2} \, \,  \overset{\fsp_{M}}{1 }  \, \, \overset{\fsu_{M}}{2} \, .
\end{align}
It is straightforward to check that the number of neutral singlets is unchanged upon freezing.

\paragraph{$\boldsymbol{\mathcal{G}_M (D_4, D_7)}$  models.}
Similarly, we can fuse a $D_4$ and a $D_7$-type theory, which results in the SUGRA quiver  
\begin{align}
\label{eq:D4D7UFquiver}
  \overset{\fsp_{3-M}}{1} \, \,  \overset{ \overset{\fsp_{3-M}}{ 1}}{ \underset{\overset{\fsp_{3-M}}{1}}{\overset{\fso_{28-4M}}{4}}}  \, \, \overset{\fsp_{11-M}}{1}  
  \, \, \textcolor{blue}{ \overset{\fso_{32}}{4} }
  \, \, \overset{\fsp_{13+M}}{1} \, \,  \overset{ \overset{\fsp_{5+M}}{ 1}}{{\overset{\fso_{36+4M}}{4}}}
    \, \, \overset{\fsp_{12+2M}}{1} \,  \overset{\fso_{24+4M}}{4} \,.
\end{align}
As argued before, the minimal number of instantons required for a $D$-type singularity increases with its rank. This in turn constrains the relative instanton numbers to lie in the range $M=1\ldots3$. 
The above model has a toric construction of threefolds $X_{D_4,D_7,M}$ via the family of polytopes $\Delta_{D_4,D_7,M}$ given in Appendix~\ref{app:ToricData}. Their respective Hodge numbers are
\begin{align}
(h^{1,1},h^{2,1})(X_{M})=(122+6M,8) \, .
\end{align} 
Freezing the model again proceeds as before. We fix $M$ to be even and exchange the respective quiver constituents to obtain
\begin{align}
\eqref{eq:D4D7UFquiver} \quad  \xrightarrow{\text{freeze }} \quad 
\overset{\fsu_{3-M}}{2}\, \, \overset{\fsp_{3-M}}{1} \, \, \overset{\fsu_{11-M}}{2} \, \,\textcolor{blue}{
\overset{\fsp_8}{1}
}\, \, 
   \overset{\fsu_{13+M} }{2}\, \, 
 \overset{ \displaystyle \overset{\fsu_{2+2M}}{1}}{  \overset{\fsu_{10+2M}}{2} }
   \, \,  \overset{\fsu_{5+M}}{2} 
\end{align}

\paragraph{$\boldsymbol{\mathcal{G}_M (\emptyset, E_6)}$ models.} 
As a final example we discuss two theories with $E_6$ type of singularities. The simplest case is fusing the $E_6$, with a trivial theory, which results in
\begin{align}  {\overset{\mathfrak{\fsp}_{16-M}}{0}}\, \, 
\textcolor{blue}{\overset{\fso_{32}}{4}} \, \, {\overset{\mathfrak{\fsp}_{M+8}}{1}} \, {\overset{\fso_{4M+16}}{4}}\,  {\overset{\mathfrak{\fsp}_{3M}}{1}} \, {\overset{\mathfrak{\fsu}_{4M}}{2}} \, {\overset{\mathfrak{\fsu}_{2M}}{2}} \, .
\end{align}
The underlying threefolds $X_M$ has Hodge numbers
\begin{align}
(h^{1,1},h^{2,1})(X_M)=(47+9M,26+M) \, .
\end{align} 
Note that the $M$-dependence of complex structure moduli comes from the non-polynomial deformations, which in turn arise from the anti-symmetric irreps of $\fsp_{16-M}$. The frozen version of the model, $\mathcal{G}^{UF}(\emptyset,E_6)_M$ is
\begin{align}  {\overset{\mathfrak{\fsp}_{16-M}}{0}}\, \, 
\textcolor{blue}{\overset{\fso_{32}}{4}} \, \, {\overset{\mathfrak{\fsp}_{M+8}}{1}} \, {\overset{\fso_{4M+16}}{4}}\,  {\overset{\mathfrak{\fsp}_{3M}}{1}} \, {\overset{\mathfrak{\fsu}_{4M}}{2}} \, {\overset{\mathfrak{\fsu}_{2M}}{2}} \, \quad\xrightarrow{\text{freeze}} \quad 
{\overset{\mathfrak{\fsp}_{8-M/2}}{0}}\, \, 
\textcolor{blue}{\overset{\fsp_{8}}{1}} 
\, \, {\overset{\mathfrak{\fsu}_{M+16}}{2}} \, {\overset{\fso_{2M+16}}{2}}\,  {\overset{\mathfrak{\fsp}_{\frac32M+8}}{1}} \, {\overset{\mathfrak{\fso}_{2M+16}}{4}} \, {\overset{\mathfrak{\fsp}_{\frac12M}}{1}} \, 
\end{align}
where the relative instanton number $M$ is again required to be even.

\section{Gauge Group Topology in the Frozen Phase}
\label{sec:1FormSyms}
In this section we comment on the global gauge group structure before and after freezing. Global properties, such as the topology of the gauge group $G$
\begin{align}
\pi_1(G) = \mathcal{Z} \, ,
\end{align}
have recently gained a lot of attention when reformulated in terms of higher form symmetries~\cite{Gaiotto:2014kfa}. Indeed, from the 10D string heterotic perspective, these features were exactly the reason why we could switch off the vector structure \cite{Witten:1997bs}. 
 
In terms of the F-theory geometry, the gauge group topology is encoded in the finite part of Mordell-Weil group $MW_{\text{tor}}(X_3)=\mathcal{Z}$  of the elliptic fibration \cite{Aspinwall:1998xj}. Note that we focus on theories without Abelian gauge factors that could mix with the centers of the non-Abelian ones and contribute non-trivially to $\pi_1(G)$ \cite{Cvetic:2017epq}. We expect our general considerations to hold also in those cases but leave a thorough exploration for future work. Since swapping to the frozen phase merely corresponds to a re-interpretation of the singularity structure of an elliptic threefold $X_3$, it does not change the MW group of $X_3$. Therefore, one expects the global 1-form gauge symmetry $\mathcal{Z}$ to be preserved upon freezing. We will focus on these aspects in more generality in Section~\ref{sec:Bounds1forms}, where we will argue for bounds on $\mathcal{Z}$ in 6D frozen F-theory vacua. 
 
Independent of whether or not a theory is coupled to gravity, there are two types of conditions for center 1-form symmetries  $\mathcal{Z}$ to be consistent. We will summarize these conditions \cite{Cvetic:2020kuw,Apruzzi:2020zot,BenettiGenolini:2020doj,Cvetic:2021sxm,Heckman:2022suy} in the following and apply them to our cases in Section~\ref{sec:1formUnfrozen}-\ref{sec:1formFrozenSUGRA} following the exposition in \cite{Heckman:2022suy}, to which we also refer for more details. 

In six dimensions, the anomaly eight-form $I_8$ factors into an anomaly four-form $I_4^i$ that consists of field strengths of gauge symmetries $G_j$ and flavor symmetries $[G_m]$ couple to the $i^\text{th}$ 2-form tensor field,
\begin{align}
I^i_{(4)} \supset  - \sum_i \eta^{ij} c_2(F_j) - A^{im} c_2 ([F_m]) \cdots \, ,
\end{align}
For the LST case, we can also have flavor symmetries. Their couplings are encoded in the coupling matrix $A^{im}$, whose entries are $1$ if the $i^\text{th}$ 2-form field is coupled to the flavor algebra $[G_m]$ and $0$ otherwise. First consider the group
\begin{align}
\tilde{G}=[\hat{G}_F] \times \prod_i^n \hat{G}_i \, ,
\end{align}
which we take to be fully non-Abelian. These factors have center symmetries 
\begin{align}
Z(\tilde{G})= Z(\hat{G}_F) \times  \prod_i Z(\hat{G}_i) \, ,
\end{align}
which are finite cyclic and Abelian groups summarized in Table~\ref{tab:Groupalpha} for all simple groups. The center symmetry $Z(\tilde{G})$ is in general broken, but there may be a linear combination $\mathcal{Z} \subset Z(\tilde{G})$ that survives and acts diagonally, such that the global structure of the Gauge group is
 \begin{align}
G=\frac{\tilde{G}}{\mathcal{Z}}=\frac{ [\hat{G}_F] \times \prod_{i=1}^{n} \hat{G}_i}{\mathcal{Z}}\, .
\end{align}
A non-trivial $\mathcal{Z} \supset Z(\tilde G)$ allows to switch on a center-twisted bundle, which is captured by the generalized Stiefel-Whitney (SW) class.\footnote{Formally, the generalized Stiefel-Whitney class in 6D is defined as $\omega \in H^2(M_6,\mathbb{Z}(\tilde{G}))$, which is the characteristic class that measures the obstruction to lift a $\tilde{G}/\mathbb{Z}(\tilde{G})$ bundle on any six-manifold $M_6$ to its cover $\tilde{G}$.} The center-twisted bundle leads to fractionalizations of the field strengths,
\begin{align}
c_2(F)= - \alpha_G \, \,  \omega (F) \wedge \omega(F) \, ,
\end{align} 
where $\omega(F)$ is the generalized SW class and $\alpha_G$ are some fractional coefficients summarized in Table~\ref{tab:Groupalpha}. Each gauge group center $Z(\hat{G}_j)$ and flavor group center $Z([G_F])$ contributes an independent SW class $\omega_j$ and $\omega_F$. The group $Spin(4N)$ is special since it has two center factors and hence, the second Chern class has two possible SW classes $w_1$ and $w_2$ that contribute as
\begin{align}
c_2(F_{Spin(4N)})=-  \frac{N}{4}(w_1 + w_2  )^2-\frac12 w_1 \cup w_2 \,.
\end{align}

The fractional coefficient $\alpha_G$ can lead to an obstruction (see~\eqref{eq:CenterAnomaly}) when performing large gauge transformation of the 2-form tensor fields which yields a non-vanishing phase in the path integral. However, one may be able to define a consistent quotient $\mathcal{Z}$, generated by a background $\omega$
which is a linear combination of center gauge twists $k^i \omega_i$ and flavor twists $k^F \omega_F$ in $Z(\tilde{G})$. Here, the $k^i$ and $k^F$ are integers that give the order of the twist element and are taken such that 2-form gauge transformations are single-valued. We collect all these twist integers in the twist vector $\vec{k}:= (k^i, k^F)$.
   
In summary, for a $\mathcal{Z} = \mathbbm{Z}_N$ symmetry to be present, we require the following two conditions be satisfied:
\begin{enumerate}
\item $\mathcal{Z}$ must act trivially on all matter representations. Each matter representation $\mathbf{R}_i$ under $\hat{G}_i$ admits a discrete center charge $q_Z \in Z( \tilde{G} )$. All matter representation of $G$ must have trivial 
  $q_\mathcal{Z}$ charge defined as 
\begin{align}
\label{eq:ChargelessMattter}
\sum_{i=1}^{n} k_i \,  q_{ Z(\hat{G}_i) }(\mathbf{R}_i  )= 0 \, \text{ mod } N \, .
\end{align}
 \item Integrality of large 2-form gauge field transformations with non-trivial SW classes must be well defined
 \begin{align} 
\label{eq:CenterAnomaly}
\sum_m  \alpha_{[G_{F,m}]} k_{F,m}^2 A^{im}  +    \sum_{\hat{G}_i}  k_j^2 \alpha_{G_j} \eta^{ij}     = 0 \text{ mod 1} \, .
\end{align}

\end{enumerate}
\begin{table}[t!]
\begin{center}
\begin{align*}
\begin{array}{|c|c|c|}\hline
G & Z(G) & \alpha_{G} \\ \hline
SU(n) & \mathbbm{Z}_n & \frac{n-1}{2n} \\ \hline
Sp(n) & \mathbbm{Z}_2 & \frac{n}{4} \\ \hline
\textit{Spin}(2n+1)& \mathbbm{Z}_2 & \frac12 \\ \hline 
\textit{Spin}(4n+2)& \mathbbm{Z}_4 & \frac{2n+1}{4} \\ \hline 
\textit{Spin}(4n)& \mathbbm{Z}_2\times \mathbbm{Z}_2 & (\frac{n}{4},\frac{1}{2})\\ \hline 
E_6 & \mathbbm{Z}_3 & \frac23 \\ \hline
E_7 & \mathbbm{Z}_2 & \frac34 \\ \hline
E_8 & \mathbbm{Z}_1 & - \\ \hline
\end{array}
\end{align*}
\end{center}
\caption{
\label{tab:Groupalpha}
Summary of centers $Z$ and fractional 
 $\alpha$ factors for any simply laced lie group $G$.
}
\end{table}
The second condition, can also be viewed as the absence of fractional BPS instantons, obtained from a $D3$ brane that wraps/intersects that is intersected
by a $D7$ brane, with non-trivial center-twist. Note that in the LST case, there is a unique linear combination
of the string charge lattice $\eta^{ij}$, which is intersected by the 6D flavor factors only leading to the condition
\begin{align}
\sum_m \alpha_{[G_{F,m}]} k^2_{F,m} \in \mathbbm{Z} \, .
\end{align}
The above condition is in fact nothing but the concistency condition for 1-form center symmetries of 8D SUGRA theories \cite{Cvetic:2020kuw}, of the gauge group $G=G_F/\mathbbm{Z}$.
\subsection{Global Structure of Unfrozen LSTs}
\label{sec:1formUnfrozen}
Before turning to compact examples, we discuss the global structure of the unfrozen vacua. As the $\textit{Spin}(32)/\mathbbm{Z}_2$ heterotic string has a non-trivial global structure in 10D, the same is true for the LST theories in six dimensions where the heterotic gauge group becomes a flavor symmetry. Indeed, the various $\textit{Spin}(32)/\mathbbm{Z}_2$ orbifold theories $\mathcal{K}_M (G)$ have a global symmetry group of type
\begin{align}
G=\frac{ [G_F] \times \prod_{m_i} Sp(k_1) \times \prod_{m_j} SO( 4 k_j ) \prod_{m_l} SU(2 k_l)          }{\mathbbm{Z}_2} \, .
\end{align}
All flavor and gauge algebra factors admit (at least one) $\mathbbm{Z}_2$ center factor on which the diagonal $\mathcal{Z}=\mathbbm{Z}_2$ quotient acts.

Before checking consistency of the $\mathbbm{Z}_2$ symmetry, we need to clarify the phases of the relevant matter multiplets under the respective centers. For $\textit{Spin}(4N)$, the vector, spinor and cospinor multiplets transform with the phases
\begin{align}
\begin{array}{c|c}
\mathbf{R} & \phi(\mathbbm{Z}_2^1,\mathbbm{Z}_2^1) \\ \hline
V& (-1, -1)  \\ 
S& (-1, 1) \\ 
C& (1,-1)\\
\end{array} \, .
\end{align} 
For $Sp(N)$, the fundamental transforms with a $-1$ phase, while anti-symmetric irreps are uncharged. The center charges $q_Z(\mathbf{R})$ for $SU(N)$ fundamentals is given by 
 $q_Z=e^{\frac{2\pi i}{N}}$, while the $m-$times (anti)-symmetric irreps have charges $q_Z=e^{\frac{2\pi m i}{N}}$. The total phase $\Phi$ for a twist  $k$ is then given by 
 \begin{align}
 \Phi= q_Z(\mathbf{R}) k \,.
 \end{align} 
This is relevant because the $Z=\mathbbm{Z}_N$ symmetry may be non-prime with $N=M T$.
In such cases, we can twist by a $\mathbbm{Z}_M$ sub-center only when $k=T$. 

In the following we discuss some theories in more detail. First, there is the  
$\mathcal{K}_M(\emptyset)$ theory with quiver
\begin{align}
[\fso_{32}] \overset{\fsp_M}{0} 
\end{align}
Evidently, there is an unobstructed twist assignment given by $k_{\fso_{32}}=(1,0)$ and $k_{\fsp_M}=1$. 

The $\mathbbm{Z}_2$ center charge contribution of the bifundamentals is
\begin{align}
  \vec{k} \cdot q_{\mathbbm{Z}_2} ( \mathbf{32}, \mathbf{2N})= k^i_{\fso_{32}} 1 +  k_{\fsp_{M}} 1 = 0\quad \text{mod}\quad 2 \, .   
\end{align}
The antisymmetric representation of $Sp(M)$ has trivial $\mathbbm{Z}_2$ center charge. 

Next we discuss the $\mathcal{K}_M(A_{2K-1})$ theories. The $SU(2K)$ gauge factors have a $\mathbbm{Z}_{2K}$ center, but only a $\mathbbm{Z}_2$ sub-center twist is activated upon choosing a twist $k=K$. The quiver (including the unobstructed twist assignments $\vec{k}$ below the quiver nodes) reads
\begin{align}
\begin{array}{lccccccc}
 &[\fso_{32}]& \overset{\fsp_N}{1} & \,\, \overset{\fsu_{2N-8}}{2}  &   \overset{\fsu_{2N-16}}{2}& \ldots  & \overset{\fsu_{2N-8(m-1)}}{2} &\overset{\fsp_{N-4m}}{1} \\[4pt] 
\vec{k}& (1,0) & (1)& (N-4)& (N-8) & \ldots & (N-4m-4)& (1)
\end{array} \, .
\end{align}
The first $\fsp_N$ gauge algebra factor leads to the condition 
\begin{align}
-1 \cdot  \frac{N}{4} + \frac{(2 N - 8 - 1)}{2 (2 N - 8)} \cdot (N - 4)^2 = \frac12( N-6) ( N-3)  \in \mathbbm{Z} \, ,
\end{align}
which is satisfied for any $N$ since either the first or the second factor is even. We can proceed similarly for the first $-2$ curve, for which we obtain
\begin{align}
\frac{N}{4} -2  \frac{2 N - 8 - 1}{2  (2 N - 8)} (N - 4)^2 + \frac{2 N - 16 - 1}{2 (2 N - 16)} (N - 8)^2  = 16 + \frac12 N (N-1) \in \mathbbm{Z} 
\end{align}
which is also always satisfied. 

Finally, we consider a $D_{4n}$ theory. For simplicity, we just choose $n=1$. The respective quiver and twist are given by 
\begin{align}
[\fso_{32}] \overset{\fsp_{8+M}}{1} \, \,  \overset{ \displaystyle \overset{\fsp_M}{ 1}}{ \underset{\displaystyle \overset{\fsp_M}{1}}{\overset{\fso_{16+4M}}{4}}}  \, \, \overset{\fsp_{M}}{1}  \qquad \text{ with }\quad \vec{k}=(  (1,0), (1), \overset{\displaystyle (1)}{\underset{\displaystyle (1)}{(1,0)}}, (1) )
\end{align}
where we have given the twist embeddings next to it. All matter representations are $\mathcal{Z}$-invariant, and with the above twists we can check the second condition~\eqref{eq:CenterAnomaly} is satisfied as well.

\subsection{Frozen LSTs}
\label{sec:1formFrozenLST}
Next, we freeze the above LSTs and show, that their global structure is preserved. 
 
We start by considering the frozen model $\mathcal{K}_N^F(\emptyset)$ given by 
\begin{align}
[\fsp_8] \overset{\fsp_{M}}{0} \, ,
\end{align}
In this case, both conditions are satisfied trivially. 

Similarly, we can take for example the $\mathcal{K}_N^{F}(A_3)$ theory
\begin{align}
\begin{array}{lccccccc}
&[\fsp_8]& \overset{\fsu_{2N}}{1} &  \overset{\fsu_{2N-8}}{1}  \\[4pt] \vec{k}& (1) & (N)& (N-4) 
\end{array}
\end{align}
All matter multiplets are neutral under the diagonal $\mathbbm{Z}_2$ action. For the two tensor multiplets, the two obstructions read 
\begin{align}
\frac{8}{4} - \frac{2N-1}{4N} (N)^2 + \frac{2N-9}{2 (2N-8)} (N-4)^2&=11-4N \in \mathbbm{Z} \,,\\  
\frac{2N-1}{4N} (N)^2 - \frac{2N-9}{2 (2N-8)} (N-4)^2&=4N-9 \in \mathbbm{Z}\,,
\end{align} 
and are satisfied for all $N$.

As a final example, we consider the $\mathcal{K}^F_{M}(D_4)$ theory with quiver
\begin{align}
\begin{array}{ccccc} 
&[\fsp_8] & \overset{\fsu_{8+2N}}{2} & \overset{\fsp_{2N}}{1 } & \overset{\fsu_{2N}}{2}  \\
\vec{k} & (1) & (N+4) & (1) & (N) 
\end{array}
\end{align} 
Again, all hypermultiplets are invariant under a diagonal $\mathbbm{Z}_2$ center action. For example, the second tensor multiplet gives the condition
\begin{align}
-\frac{2 N}{4} +  \frac{2 N + 8 - 1}{2 (2 N + 8)}(N + 4)^2 + \frac{2 N - 1}{2 (2 N)} (N)^2 = 7 + 3 N + N^2 \in \mathbbm{Z} \,,
\end{align}
which are again satisfied automatically.

\subsection{Fusing LSTs to obtain SUGRA theories}
\label{sec:1formFrozenSUGRA}
In Section~\ref{sec:frozengravity} we have discussed how LSTs can be fused to obtain SUGRA theories. Each constituent in the fusion has a non-trivial center symmetry, which we now show to be invariant under fusion. In order to do so, we simply need to check that the fusion nodes $ \overset{\fso_{32}}{4}$ in the unfrozen case and $ \overset{\fsp_{8}}{1}$ in the frozen case leaves the condition~\eqref{eq:CenterAnomaly} invariant. First recall that there must be 24 instantons in total on the $\fso_{32}$ node, and hence 
\begin{align}
 \ldots  \overset{\fsp_{24-N}}{1} \,  \overset{\fso_{32}}{4} \,  \overset{\fsp_{N}}{1} \ldots  
\end{align}
The consistency condition results in
\begin{align}
 -4 \frac{8}{4} + \frac{24-N }{4}+\frac{N}{4}=16 \in \mathbbm{Z}\, ,
\end{align}
One can proceed similarly for the frozen phase. Here, the $\fsp_8$ gauge node can connect to either $\fsu$ or $\fsp$ gauge algebra factors. Anomaly cancellation fixes such configurations locally to one of the three cases 
 \begin{align}
 1.\quad [ \fsp_{12-M}] \overset{\fsp_8}{1}  [\fsp_M] \, , \qquad
 2. \quad [ \fsp_{24-2M}] \overset{\fsp_8}{1}  [\fsp_{M}] \, , \qquad
 3. \quad [\fsu_{24-2M}] \overset{\fsp_8}{1}  [\fsu_{2M}]\, ,
 \end{align} 
with $M=2N$. The twist for the $\fsu_{2K}$ factors is again fixed by the $\mathbbm{Z}_2$ sub-center to be $k=K$. With this choice, all obstructions evaluate to integer classes and hence the $\mathbbm{Z}_2$ 1-form gauge symmetry is preserved in the SUGRA theory.

\subsection{Bounds on the 1-form symmetry sector}
\label{sec:Bounds1forms}
The authors of \cite{Lee:2022swr} consider general conditions on the 1-form gauge symmetry sector $\mathcal{Z}$ in 6D SUGRAs with only non-Abelian gauge symmetry factors. Note that any theory with $T>0$ tensor multiplets has a heterotic string in its spectrum. In terms of the F-theory geometry, this heterotic string is obtained from a $D3$ brane that wraps a $\mathcal{C}^2=0$ curve in the base. As the gauge group topology $\mathcal{Z}$ is a global property, it must be consistent with the (perturbative) heterotic gauge symmetry $G_{Het}$. The latter is fixed by the finite choice of possible 8D SUGRA theories, consistent with the fact that the geometry is an elliptic K3 fibration. Thus, the 6D 1-form symmetries are bounded by the possible of 8D SUGRA vacua. This is consistent with the fact that the 1-form anomaly in 8D coincides with the one found on the respective 0-curve in 6D \cite{Cvetic:2020kuw}. In summary, the 1-form gauge symmetry in 6D must take on one of the following possible values 
 \begin{align}
 \label{eq:Z1Bounds}
\mathcal{Z}^{UF}= \mathbbm{Z}_{n} \times \mathbbm{Z}_m \text{ with } (n,m) \in \{ (1,\{1,2,3,4,5,6 \}), (2,\{2, 4\}),(3,3)\} \, ,
 \end{align}
which are all the allowed 8D values with the exception of $\mathbbm{Z}_7,\mathbbm{Z}_8$ and $\mathbbm{Z}^2_4$. 

We want to extend these considerations to frozen 6D F-theory vacua and show that these do not contribute any new factors, i.e., the list \eqref{eq:Z1Bounds} is complete. In fact, we argue that frozen 6D F-theory vacua are very strongly constrained, such that their 1-form center gauge symmetries can only be
\begin{align}
\label{eq:Z1BoundsFr}
\mathcal{Z}^{F}= \{ \mathbbm{Z}_2, \mathbbm{Z}_2 \times \mathbbm{Z}_2 \} \, .
\end{align}
When considering a 6D F-theory SUGRA theory with $T>0$ and a single $O7^+$ brane, there is again a heterotic string in the spectrum, which lives on the $0$ curve. 
However, in the frozen phase, the local theory of this heterotic string should correspond to the $\textit{Spin}(32)/\mathbbm{Z}_2$ vacuum without vector structure.\footnote{Note that 8D CHL vacua live in the same component of the moduli space as those without vector structure \cite{Lerche:1997rr}.} Thus, running a similar argument as above, we simply check the consistent 8D CHL or $\textit{Spin}(32)/\mathbbm{Z}_2$ 1-form gauge symmetries, which correspond to the constraints for the possible 6D 1-form symmetries $\mathcal{Z}$. The latter have been explored recently in \cite{Cvetic:2021sjm,Font:2021uyw} and (in the absence of Abelian gauge factors) where shown to be the ones of~\eqref{eq:Z1BoundsFr}. 

From the perspective of the frozen F-theory geometry, the above statement may also be seen the following way: Recall that freezing in F-theory amounts to an \textit{re-interpretation} of the singularity structure in the elliptic fiber: An $I_4^*$ fiber, which is interpreted as an unfrozen $\fso_{16}$ gauge algebra on an
$8D7+O7^-$ stack gets swapped to a frozen $O7^+$ plane with no gauge algebra in the frozen phase. The global gauge group structure, on the other hand, is encoded geometrically in the Mordell-Weil torsion group, which stays invariant under freezing. But Mordell-Weil torsion restricts the possible SL$(2,\mathbbm{Z})$ axio-dilaton monodromies to lie in an $\Gamma_1 (n)$ or $\Gamma(m)$ sub-group for $\mathcal{Z}=\mathbbm{Z}_n$ or $\mathcal{Z}=\mathbbm{Z}_{m}^2$ MW groups, respectively~\cite{Hajouji:2019vxs}.\footnote{For $\mathbbm{Z}_2 \times \mathbbm{Z}_4$, monodromies $M$ are constrained to be in a non-classical congruence subgroup $M \in  \Gamma_1(4) \cap \Gamma(2)$.} Recall that the monodromy of an $I_4^*$ fiber coincides with that of an $O7^+$,
\begin{align}
M(I_4^*)=M(O7^+) = \left( \begin{array}{cc} 
-1 & 4 \\ 
0 & -1 \\
\end{array} 
     \right) \ni \Gamma(2) \, .
\end{align}
Therefore, the maximal SL$(2,\mathbbm{Z})$ congruence sub-group a single $O7^+$ plane can leave invariant is $\Gamma(2)$. which corresponds to a $\mathbbm{Z}_2 \times \mathbbm{Z}_2$ Mordell-Weil torsion group. This in turn is nothing but the statement that the group $\textit{Spin}(16)$ has a $\mathbbm{Z}_2 \times \mathbbm{Z}_2$ center that we may be able to gauge.

A related geometric argument has been given in \cite{Bershadsky:1998vn}, where frozen 8D F-theory vacua where interpreted as switching on a discrete $\mathbbm{Z}_2$ background flux for  $B_2$ in the $\mathbbm{P}^1$ base. As $B_2$ and $C_2$ form an SL$(2,\mathbbm{Z}_2)$ doublet, such a discrete flux background is only well-defined when the monodromies $M$ are restricted to 
\begin{align}
M  \in \Gamma_0(2) \, .
\end{align}
In order to admit a non-trivial global structure, such as $\mathcal{Z}=\mathbbm{Z}_n$ or $\mathbbm{Z}_m^2$ , we must then have monodromies that preserve $\Gamma_1(n)$ or $\Gamma(m)$ subgroups of $\Gamma_0(2)$. These subgroups are
\begin{align}
\Gamma_2 \subset \Gamma_1(2) \subset \Gamma_0(2) \, ,
\end{align}
and thus only the groups \eqref{eq:Z1BoundsFr} are consistent. Note, however, that also a $\Gamma_1(4)$ group, i.e., a $\mathbbm{Z}_4$ symmetry, can be possible.

Of course, one may wonder whether there exist other exotic types of non-perturbative $O7$ orientifold planes that may be of order three or higher that may be compatible with a higher order global structures. This is not the case, as argued in \cite{Tachikawa:2015wka}.

\section{Conclusions}
\label{sec:Conclusions}
In this work we have extended and explored the geometric dictionary of frozen F-theory vacua in six dimensions. We explored the frozen phase by constructing F-theory duals of $\textit{Spin}(32)/\mathbbm{Z}_2$ heterotic NS5 brane theories, and their orbifolds with and without vector structure for all ADE groups. We show that turning off vector structure corresponds to a folding of the respective quiver diagram by a $\mathbbm{Z}_2$ outer automorphism, which freezes a large amount of tensor multiplets. 

We explain how freezing can be implemented by a change of a divisor in the base of the F-theory threefold and show consistency with rules established in \cite{Bhardwaj:2018jgp}. Since the theories are heterotic little string theories, we also establish \textit{fusion rules} that allow to combine them into consistent supergravity backgrounds, which we construct systematically in toric geometry. From this construction, we show that freezing does not change the number of neutral hypermultiplets in the theory.

Finally, we comment on the global gauge symmetry structure in frozen models, i.e., the discrete center 1-form gauge symmetry sector. This sector is strongly constrained and lies within the bounds given in \cite{Lee:2022swr}. 

There are various open questions which we would like to come back to in future works. In particular, it is natural to ask whether we can study orbifolds of NS5 branes in heterotic CHL orbifolds in a similar fashion. Unfortunately, there are no perturbative methods to do so directly, but these theories could be T-dual to our $\textit{Spin}(32)/\mathbbm{Z}_2$ instanton models without vector structure. Furthermore ,it would be exciting to relate the construction of frozen 8D F-theory vacua with a discrete 2-form background field~\cite{Bershadsky:1998vn} and the 6D constructions explored in this work. We have already seen that 8D frozen F-theory vacua are necessary to understand the disconnected components of the moduli space. It would therefore be very interesting to understand whether 6D (or lower-dimensional) frozen F-theory vacua may also correspond to disconnected components of the SUGRA moduli space, or whether there (always) exists some dynamical transition from an unfrozen component. Moreover, it would be interesting to study this relatively little explored framework in the context of string model building. Reducing gauge group ranks and freezing moduli by exploiting $O7^+$ planes could be promising for moduli stabilization and to circumvent the tadpole problem~\cite{Bena:2020xrh}. 

Finally, our detailed investigation of the duality between the $\textit{Spin}(32)/\mathbbm{Z}_2$ heterotic string without vector structure and the frozen phase of F-theory revealed a number of new constraints and examples in studying compact models with frozen singularities. In one direction, it would be interesting to develop a systematic procedure to construct $6$d F-theory compactifications with $O7^+$-planes and to analyze more subtle aspects of their physics, such as the global forms of the gauge groups, discrete symmetries, and abelian factors. In another direction, it would also be desirable to make precise the relations between the appearance of $O7^+$-planes, the absence of vector structure, $B$-fields and discrete torsion, as described in~\cite{Kakushadze:1998bw,Kakushadze:1998cd,Kakushadze:2000hm}.

\subsection*{Acknowledgments}
The authors thank Yuji Tachikawa, Alessandro Tomasiello and David Morrison for helpful discussions. The work of FR is supported by the NSF grants PHY-2210333 and PHY-2019786 (The NSF AI Institute for Artificial Intelligence and Fundamental Interactions). The work of FR and PKO is also supported by startup funding from Northeastern University. PKO and BS would like to thank the KITP and the program "What is String Theory?  Weaving Perspectives Together" during the completion of this work. This research was supported in part by grant NSF PHY-2309135 to the Kavli Institute for Theoretical Physics (KITP).

\appendix
\section{Toric Data of elliptic threefolds}
\label{app:ToricData}
In this appendix, we summarize the toric vertices for two families of threefolds used in the examples in Section~\ref{sec:frozengravity}. We give the families of vertices of the 4D polytopes $\Delta$ from which the respective threefolds can be constructed following Batyrev. The vertices also allow us to compute the Hodge numbers. The toric rays are
\begin{align}
\Delta_{(\emptyset,D_4,M)}=\left\{
\begin{array}{ c }
(0,1,0,0)\\
(1,0,0,0)\\
(0,-1,0,1)\\
(0,-1,M,1)\\  
(-2,-3,-2,2)\\
(-2,-3, -8,-1)\\
(-2,-3,8-M,-1)\\
(-2,-3,2M+2,2)\\
\end{array}  \right\}\, , \qquad
\Delta_{(D_4,D_7,M)}=\left\{
\begin{array}{c} 
(0,1,0,0)\\
(1,0,0,0)\\
(1,0,-2,2)\\
(1,1,0,1)\\
(-2,-3,-8,2)\\
(1,0,2M,2)\\
(-2,-3,6+2M,2)\\
(1,1,3-M,1)\\
(-2,-3,-4,-2)\\  
(0,-1,-1,-1)\\
(-2,-3, 6-2M,-2)\\
(0,-1,-1,-1)\\
(0,-1,2-M,-1)
\end{array}
\right\}\,.
\end{align}

\bibliographystyle{JHEP}
\bibliography{refs}

\end{document}